\documentclass[aps,twocolumn,pra,amssymb,amsmath,floatfix,superscriptaddress,nofootinbib,float,romannum]{revtex4}

\usepackage[dvipsnames,usenames]{xcolor}
\usepackage{graphicx}
\usepackage{mathptmx}
\usepackage{bm}
\usepackage{setspace} 

\usepackage[a4paper, total={18cm, 24cm}]{geometry}

\newcommand{\ophi}{\overline  \varphi}
\newcommand{\os}{\overline  s}
\newcommand{\ot}{\overline  t}
\newcommand{\ou}{\overline  u}

\newcommand{\OV }{\overline V}

\begin{document}
\title{Conservation, crossing symmetry,
and completeness in diagrammatic theories}
 
\author{Frederick Green}
\email{frederickgreen@optusnet.com.au}
\affiliation{
School of Physics, The University of New South Wales,
Sydney, NSW 2052, Australia}

\begin{abstract}
The diagrammatic analysis of interacting particle assemblies harbors a
fundamental mismatch between two of its main implementations:
$\Phi$-derivable (conserving) approximations and parquet (crossing symmetric)
models. No termwise expansion, short of the exact theory itself, can
be both conserving and crossing symmetric. This work applies the Kraichnan
embedded-Hamiltonian formalism for strongly coupled systems to investigate
consistency of the interplay between purely pair-mediated correlations and
pair-irreducible ones. The approach sheds a different light on the issue of
crossing symmetry versus conservation. In the process, the parquet equations
acquire a different formulation. 

\vskip 0.25cm
\centerline{{\bf NOTE}. Corrected redraft of
Phys. Rev. A {\bf 110}, 052222 (2024). Ver. K3\_250830} 
\end{abstract}

\maketitle

\section{Introduction}

A puzzling characteristic of many-body perturbation expansions concerns an
inherent constraint on their practical applicability to interacting systems
\cite{becker,js,roger}.
It is not possible for any approximate series to incorporate both of the
two fundamental attributes of the underlying exact description: microscopic
conservation, and the dominance of particle statistics in the correlated
state as implemented in the exchange, or crossing, symmetry of the exact
two-body scattering amplitude.

Only the full diagrammatic expansion can satisfy both conservation and
crossing symmetry. The impossibility for any subseries to do so was
discussed initially by Becker and Grosser for nuclear scattering
\cite{becker},
then more generally by Jackson and Smith
\cite{js}
and finally in great detail by Smith
\cite{roger},
who analyzed the parquet-like structure of $\Phi$-derivable two-body
scattering amplitudes in the context of the failure of crossing symmetry.

That limitation leaves just one of two mutually exclusive options for
calculation. Either one selects an interacting model that conserves
microscopically and globally but lacks crossing symmetry; or else, say in
the specific case of fermion scattering, one sets up a model that respects
antisymmetry and is richer in correlations, but nonconserving.

The incompatibility of crossing symmetry and conservation in
diagrammatics shows itself as a mismatch in the approximate treatment of
the single-particle self-energy. If derived variationally from a model
Luttinger-Ward (LW) functional
\cite{lw,kb1,kb2},
then even if the two-body scattering kernel defining the LW object is
crossing symmetric, the two-body scattering kernel subsequently derived
from the associated self-energy will carry additional noncrossing symmetric
terms that must be retained for the model's transport and dynamic response
descriptions to be fully conserving. The origin of the mismatch lies in an
ambiguity in defining the truncated LW kernel, unlike the exact object
which is evidently unique.

On the other side, if one solves the appropriate parquet equations
selfconsistently
\cite{pqt1,pqt2,pqt2a,pqt3},
a manifestly crossing symmetric two-body scattering kernel is obtained which
then defines an associated self-energy. Even as it exhibits more structure
than its closest $\Phi$-derivable analog, just as for the latter the parquet
self-energy also generates, variationally, new noncrossing symmetric
additions to its originating two-body scattering amplitude. These must be
discarded to keep crossing symmetry, but since {\em within the model} they
represent physically consistent scattering effects, removing them takes away
the variational consistency needed for conservation.

Our goal is to make more transparent what it is about diagrammatics that
underlies the seemingly odd fact, at least for fermions, by which
conservation and the natural outcome of particle statistics cannot coexist
in any rational approximation. We do not attempt to overcome a demonstrably
intractable limitation. Rather, we offer a conceptually different insight
into a many-particle system's graphical constitution and a different outlook
on the character of both $\Phi$ derivability and parquet theory.

Our instrument is the Kraichnan formalism
\cite{k1,k2}.
whose two-step construction leads to an extended Hamiltonian description for
a basic class of $\Phi$-derivable approximations, performing all-order
diagram summations abstracted in a strictly conserving fashion from the
exact system expansion. By preserving the Hamiltonian's unitarity, the
approach also safeguards the structure of all sum rules based on causality.
It has been applied to particular cases
\cite{KI}
to establish the general basis for their satisfaction of the full set of
conservation relations: for example, the dynamical sum-rule constraints
on the density-density and related correlation functions
\cite{dp,pn}.
Identities dependent upon the completeness of Fock space are not preserved
and, as will be shown, the breakdown of crossing symmetry in
$\Phi$ derivability is intimately related to this feature. 

To compare parquet and $\Phi$ derivability we start by applying Kraichnan's
Hamiltonian embedding to a familiar subtype of selfconsistent correlations:
those mediated only by the three possible avenues for purely two-body
propagation between successive interactions. They are: particle-particle
and hole-hole ladders ($s$ channel) and particle-hole polarization bubbles
($t$ channel) with the latter's exchanges, the particle-hole ladders
($u$ channel)
\cite{KII}.
This exclusively pairwise ``$stu$'' description of excitations,
here couched in very different terms, is identical computationally to the
$\Phi$-derivable theory of Scalapino and co-authors known as the fluctuation
exchange, or FLEX, model
\cite{sb}.
See also the review by Bickers
\cite{pqt3}.

Section II briefly revisits the theory of the Luttinger-Ward correlation
energy functional
\cite{lw}.
Two standard representations of the LW functional are introduced,
both parametrized by the interaction potential and, in one case,
by the renormalized one-particle Green function
\cite{lw}
while the other uses the renormalized two-particle Green function
\cite{ddm,kita}.
There follows an overview of Kraichnan's construction for the $stu$-FLEX
model, indicating significant features derived from the formalism.
In Sec. III irreducibility in the exact expansion of the LW functional is
discussed (here ``irreducible'' denotes a diagram not generated
within the $stu$ framework).
The exact parquet equations are given an alternative derivation, from which
an alternative interpretation to the standard one also follows, where
Kraichnan again provides the tool.
Section IV analyzes the loss of completeness and failure of crossing
symmetry in terms of how Kraichnan stochastics operates. There follows
a related phenomenon in the two-body description of the LW functional: the
breakdown of the identity relating static and dynamic structure factors.

We close with Sec. V summarizing our conclusions and several implications
for future study. Appendices A and B recall the accounting numerics, first
for diagrams with less than maximum topological symmetry and second
for the Kraichnan average of closed diagrams. Appendix C covers
identification of the primitively irreducible two-body scattering kernel.

\section{Essentials}

\subsection{Luttinger-Ward functional}

We recapitulate the Kraichnan formalism's objective: the structure of the
interacting ground state.  To simplify the discussion we take a spatially
homogeneous system in the momentum-spin representation $k$ at zero
temperature, although the formalism works for any two-body interaction
for uniform and nonuniform cases, at finite temperature or not
\cite{k1,k2}.

The standard system Hamiltonian comprises the one-body kinetic-energy part
determined by the reference basis states. The kinetic energy does not play
a direct role in the correlation analysis and is not considered further.
We focus on the two-body interaction operator
\cite{pn}:
\begin{eqnarray}
H_i[V]
&=&
\frac{1}{2} {\sum_{k_1 k_2 k_3 k_4}}\!\!\!\!'
{\langle k_1 k_2 | V | k_3 k_4 \rangle}~
a^{\dagger}_{k_1} a^{\dagger}_{k_2} a_{k_3} a_{k_4}
\label{irr00}
\end{eqnarray}
The sum over states has the momentum and spin
conservation restriction $k_1 + k_2 = k_3 + k_4$. We absorb a factor of
inverse system volume into $V$, the interaction potential
\cite{pn}.
Indices $k, q$, etc. may address a single-particle phase space in more
than one dimension.

Central to the development of conserving approximations and certainly to
$\Phi$-derivable models possessing an explicit Hamiltonian, is the
Luttinger-Ward functional: the component of the ground
state energy (generally, the free energy) manifesting, and in one sense
generating, its full interacting structure
\cite{lw}.
The LW functional provides the correlation energy in the ground state
$\psi_0$, expressed as a Hellmann-Feynman coupling constant integral:
\begin{eqnarray}
\Phi[V]
&\equiv&
\int^1_0 \frac{dz}{2z} {\langle \psi_0[zV] |H_i[zV]| \psi_0[zV] \rangle}.
\label{aux01}
\end{eqnarray}

There are two distinct but equivalent ways to describe the correlations in
the LW functional. The approach closer to the analysis of Luttinger and Ward,
extended to conserving approximations by Kadanoff and Baym
\cite{kb1,kb2},
addresses $\Phi$ in terms of the fully renormalized one-body propagator
and self-energy. Closer to parquet analysis in stressing two-body
processes
\cite{pqt1,pqt2,pqt2a}
is the theory of the pair correlation function in its static and dynamic
forms; more particularly, its Fourier transform, the structure factor.

\subsubsection{Pair-correlation description of $\Phi$}

The following is based on Pines and Nozi\`eres
\cite{dp}.
In a closed interacting system the static structure factor measures the
instantaneous correlation between a pair of constituent particles.
Its formal definition is
\begin{eqnarray}
S(q)
&\equiv&
N^{-1} {\left(
{\langle \psi_0|\rho^{\dagger}_q\rho_q|\psi_0 \rangle} -
{\langle \psi_0|\rho^{\dagger}_0\rho_0|\psi_0 \rangle} \delta_{0q}
\right)}
\cr
&=& N^{-1} \frac{\delta \Phi}{\delta V(-q)} - N\delta_{0q}
\label{aux02}
\end{eqnarray}
in which $\rho_q \equiv \sum_k a^{\dagger}_{k+q}a_k$ generalizes
the particle number operator
\cite{dp}
and $N = {\langle \psi_0|\rho_0|\psi_0 \rangle}$
is the total particle number.
The LW functional has expression as a two-body object:
\begin{eqnarray}
\Phi[V;S]
&=&
N {\left( \int^1_0 \!\! \frac{dz}{2z} \sum_q zV(-q) S[zV](q)
 \!+\! \frac{N}{2} V(0) \right)};
~~~ ~~ 
\label{aux02.1}
\end{eqnarray}
here we emphasize the implicit functional dependence of $S(q)$
on the interaction within the Hellmann-Feynman formula.

In the exact problem the static structure factor is also expressible
as the inverse Fourier transform, at equal times,
of the particle-number autocorrelation $S(q,\omega)$ in the frequency
domain. As the response to a weak perturbation term $\sim U\rho$
added to the Hamiltonian, the dynamic structure factor is determined by
the departure from the unperturbed state:
\begin{eqnarray}
S(q,\omega)
&\equiv&
-\frac{1}{\pi} {\rm Im}{\bigg\{
\frac{\delta(\rho[U] - \rho[0])}{\delta U(-q,-\omega)} \bigg\}},
\label{aux03}
\end{eqnarray}
satisfying the identity
\cite{dp}
\begin{eqnarray}
S(q)
&\equiv&
\frac{1}{N} \int^{\infty}_0 d\omega S(q,\omega)
\label{aux04}
\end{eqnarray}
Insofar as it holds for the exact case, the importance of Eq. (\ref{aux04})
as one nexus between the static form Eq. (\ref{aux02}) and the dynamic form
Eq. (\ref{aux03}), will become apparent in the context of a $\Phi$-derivable
approximation's departure from the exact expansion.
In conserving approximations the consistent interpretation of the two
structure factors needs care, and we postpone this to Sec. IV. Their
mismatch could act as a numerical fidelity check on such models, but its
true significance is conceptual.

We go on to specify the LW functional in terms
of the single-particle Green function, or propagator, $G$ and its associated
self-energy $\Sigma$. We follow Kadanoff and Baym
\cite{kb1,kb2}
and the original paper of Luttinger and Ward
\cite{lw}.

\subsubsection{One-body description of $\Phi$}

The original form of the exact Luttinger-Ward functional
\cite{lw}
is also a coupling-constant integral, expressed in terms of
renormalized one-body quantities:
\begin{eqnarray}
\Phi[V]
&\equiv&
\int^1_0 \frac{dz}{2z} G[zV]\!:\!\Sigma[zV;G]
\cr
&=&
\int^1_0 \frac{dz}{2z}
G[zV]\!:\!{\Bigl( \Gamma[zV; G]\!:\!G[zV] \Bigr)};
\label{irr03}
\end{eqnarray}
we explore the second right-hand-side expression shortly.
Each dot on the right-hand side of Eq. (\ref{irr03})
denotes an internal summation over momentum-energy and spin.

As usual the Dyson equation defines the one-body propagator $G$
determining the LW functional:
\begin{eqnarray}
G
&=&
G_0 + G_0\cdot \Sigma[G]\cdot G.
\label{aux05}
\end{eqnarray}
In the momentum-energy representation the noninteracting propagator
is $G_0(k,\omega) \equiv (\omega - \varepsilon_k + \mu)^{-1}$ with $\mu$
the Fermi energy.
The Dyson equation is inherently selfconsistent since the self-energy
$\Sigma$ is the variation of $\Phi$ with respect to $G$:
\begin{eqnarray}
\Sigma[G[V]]
&\equiv&
\lim_{z \to 1}{\left\{ \frac{\delta \Phi}{\delta G[zV]} \right\}}.
\label{aux05.1}
\end{eqnarray}
Note that this variation is restricted. It treats Eq. (\ref{irr03}) as
a functional of $G[zV]$, not of the full physical one-body propagator
$G[V]$. Diagrammatically it is equivalent to the full variation
$\delta\Psi/\delta G[V]$ for the dressed Luttinger-ward functional
$\Psi$ introduced in Eq. (\ref{lw47.1}) of Appendix A.
Unless noted otherwise, variations with respect to $G$
in the main text following are understood to conform to its use
in Eq. (\ref{aux05.1}).

Although the abstract properties of the correlation energy functional
$\Phi$ have a nonperturbative development, either in terms of of $S$ and $V$
\cite{kita}
or of $G$ and $\Sigma$
\cite{potthoff,lin1},
Eqs. (\ref{irr03})--(\ref{aux05.1}) would remain a computational tautology
without some understanding of the LW functional's diagrammatic provenance.
For this, we recapitulate the graphical content of its {\em exact} structure
\cite{lw,lin2}.
\begin{itemize}
\item[]
(A) The object $(\Sigma-G:V):G = S::V$ is the collection of all closed
Feynman skeleton diagrams beyond Hartree, to every order in $V$.
A skeleton diagram cannot be cut into two disjoint parts by
severing any one pair of internal lines $G$. The coupling-constant integral
Eq. (\ref{irr03}) automatically generates the required
combinatorial factors at every order of the interaction while
the diagrammatic topology is fixed by the integrand $\Sigma:G$.
\item[]
(B) $\Phi$ is invariant under particle pair exchange.
\item[]
(C) The kernel $\Gamma[V;G]$ is unique and microscopically reversible:
${\langle k_4 k_3 | \Gamma | k_2 k_1 \rangle} =
{\langle k_1 k_2 | \Gamma | k_3 k_4 \rangle}^*$
and every internal line $G$ is renormalized selfconsistently with the same
$\Sigma[V;G]$ where
\begin{eqnarray}
\Sigma[V;G] = \Gamma[V;G]:G.
\label{aux05.11}
\end{eqnarray}
The two-body scattering kernel $\Gamma$, introduced in the second line of the
definition of $\Phi$, Eq. (\ref{irr03}), is the central quantity in comparing
approximations to the correlation structure with the ideal specification of
$\Phi$.
By nature of the exact state and for the exact state alone,
$\Gamma$ carries every physically possible mode of interaction within the
system. It leads to the fourth basic
principle underlying the full LW functional:
\item[]
(D) the exact ground-state description is self-contained. No new interaction
configuration, not already manifest in the kernel $\Gamma$ itself, can arise
from the second variation of the LW functional with $G$. That is,
\begin{eqnarray}
\frac{\delta^2 \Phi[V;G]}{\delta G \delta G} = \Gamma[V;G].
\label{aux05.2}
\end{eqnarray}
\end{itemize}

Item (A) does not apply in approximation since only a subset of the LW
functional's complete diagrammatic content can ever be incorporated. While
a $\Phi$-derivable model will satisfy Eq. (\ref{aux05.1}) by construction
\cite{kb1}
and is normally expected to satisfy (B)--a notable exception being the
classic random-phase approximation
\cite{dp}--it will not satisfy (D) other than trivially (Hartree-Fock).

The physical context for Kraichnan's formalism has been set out.
Now we recall its constitution.

\subsection{Kraichnan theory}

\subsubsection{Basic conception}

\vskip -0.15cm
\begin{figure}
\centerline{
 \includegraphics[height=4.5truecm]{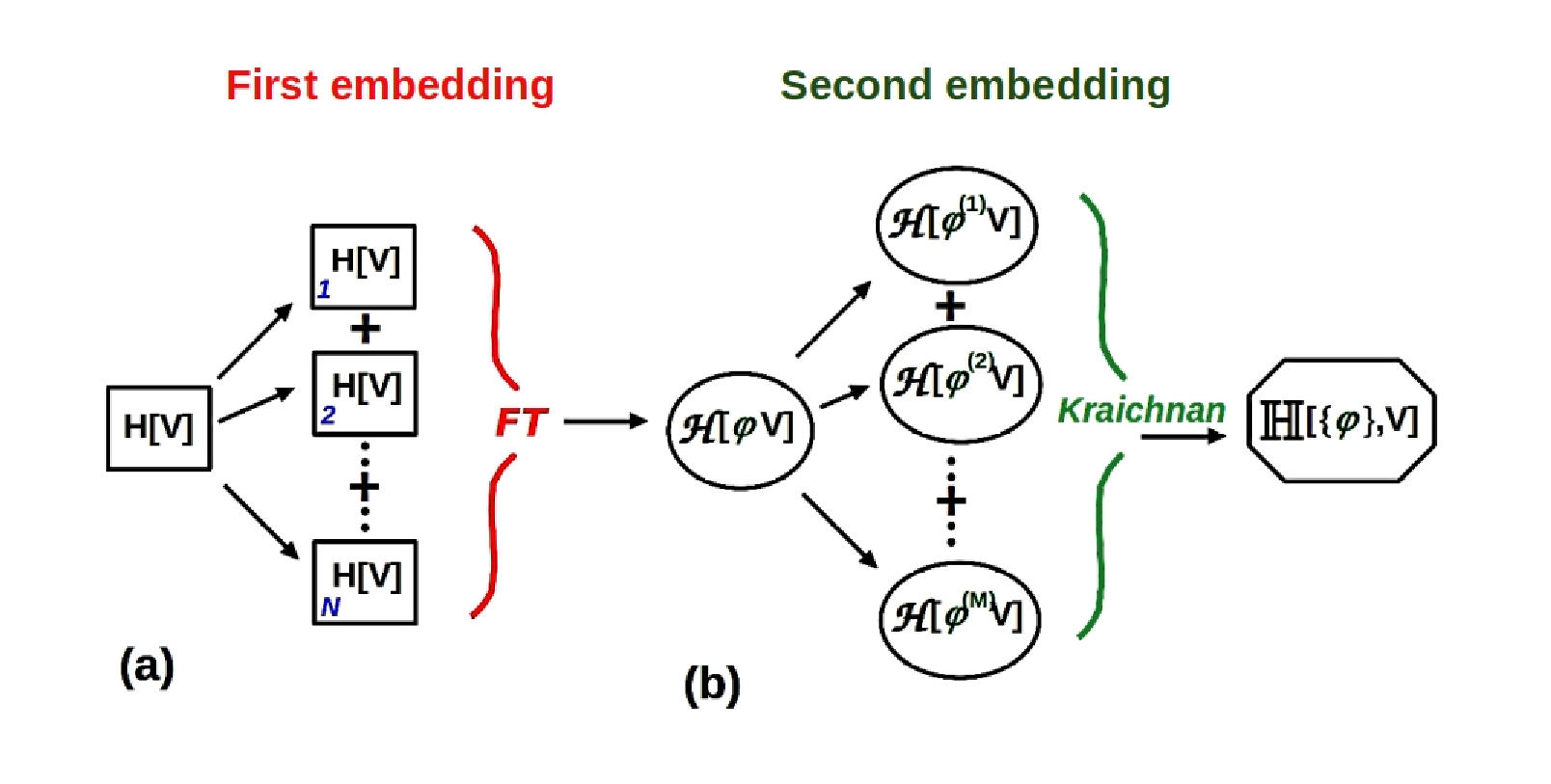}
}
\caption{
Construction of the Kraichnan Hamiltonian. (a) The exact Hamiltonian is
embedded first in an arbitrarily large sum of ${\cal N}$
identical but distinguishable copies, indexed by $n = 1,2, ...{\cal N}$.
A Fourier transform over the index generates a collective description.
The interaction ${\langle k_1k_2|\OV|k_3k_4 \rangle}$ is
augmented with a parameter $\varphi_{\nu_1\nu_2|\nu_3\nu_4}$ transforming
in its Fourier indices $\nu$ exactly as $V$ in its physical indices $k$.
(b) The collective Hamiltonian is next embedded in an arbitrarily large sum of
${\cal M}$ topologically identical replicas, except that each now carries a
unique set of factors $\varphi$. The extended ${\cal NM}$-sized
Hamiltonian remains Hermitian. Setting $\varphi$ to unity recovers the
exact physical expectations. When $\varphi$ is tailored to be randomly
assigned over the ${\cal M}$-fold ensemble of collective Hamiltonians,
a selected subset of correlation diagrams is distinguished by their
total product of coupling factors factoring out to unity. On averaging over
the stochastic distribution, random phasing suppresses everything else. All
canonical relationships valid for the underlying Hamiltonian remain valid
in the reduced model.
}
\label{F1}
\end{figure}

Figure \ref{F1} schematizes the two-step process arriving at an embedding of
the physical Hamiltonian $H$ in an extended object that can be modified while
preserving the Hermitian property of the original. Construction of this
extended Hamiltonian is complicated. We set out its various constructive
steps.

\vskip 0.15 cm
\noindent
{\em Step 1A}.\\
Take a large number ${\cal N}$ of identical but distinguishable
copies of the Hamiltonian, Eq. (\ref{irr00}). Label them with
$n = 1, 2, ... {\cal N}$. The label could be thought of as a pseudospin, but
we will consider it to satisfy periodicity just like the physical label on a
real-space unit cell within a uniform lattice; that is, $n + {\cal N}$
is identified with $n$. We let ${\cal N}$ tend to infinity in the limit.

\vskip 0.15 cm
\noindent
{\em Step 1B}.\\
A ``collective'' index $\nu$ is defined via a Fourier transform over $n$,
detailed in Eq. (\ref{astra}) following. This acts exactly like the
momentum label $k$ that collectively encodes the individual unit cells in
the real-space lattice.

\vskip 0.15 cm
\noindent
{\em Step 1C}.\\
The ${\cal N}$-fold collective Hamiltonian inherits the interaction $V$. When
this is not modified, expectations over the collective indices still recover
the original properties of any member $H$. However, a novel degree of freedom
is introduced via $\nu$ and we exploit it by attaching to
${\langle k_1 k_2 | V | k_3 k_4 \rangle}$ a new coupling factor
$\varphi_{\nu_1\nu_2|\nu_3\nu_4}$.

The symmetries of $\varphi_{\nu_1\nu_2|\nu_3\nu_4}$ in its $\nu$
indices are identical to that of ${\langle k_1 k_2 | V | k_3 k_4 \rangle}$
on its $k$ labels. The modified collective Hamiltonian
stays Hermitian, with real eigenvalues (these must differ from the exact case
except when $\varphi$ is unity).

\vskip 0.15 cm
\noindent
{\em Step 2A}.\\
The ${\cal N}$-fold collective Hamiltonian, carrying its own $\varphi$, is a
well formed entity in its own right. We now generate a new ${\cal M}$-fold
set of such replicas for ${\cal M}$ large, and sum them to create a far
bigger Hamiltonian. Here, in contrast with the first step, each Hamiltonian
member comes with a particular collection of $\varphi_{\nu_1\nu_2|\nu_3\nu_4}$
which differs from one replica to the next in the superassembly. Nevertheless
the total ``Hamiltonian of Hamiltonians'' is still Hermitian.

\vskip 0.15 cm
\noindent
{\em Step 2B}.\\
For each quadruple of collective indices $[\nu_1,\nu_2,\nu_3,\nu_4]$ the
${\cal M}$ values for the coupling $\varphi$ are then defined as functions
of randomly distributed parameters. See Eq. (\ref{irr02}) and Fig. \ref{F2}.

We are at the heart of Kraichnan's procedure. The functional form of the
couplings is specifically tailored so that, taking a double expectation
(or, Kraichnan average) over the indices and the stochastic distributions
of the parameters, only certain subsets of terms are sure to survive while
all else drops out by destructive interference among random coefficients when
these do not cancel mutually to yield an overall product of unity. (Details
of the process are in Sec. IIF and Appendix B below, notably how the
constraint on the indices operates just like conservation of momentum.)

Since the Hermitian structure is preserved in taking expectations over the
superassembly, the fundamental identities between expectation values that
depend on hermiticity still hold after Kraichnan averaging. This guarantees,
first and foremost, that the conservation laws are satisfied by the
approximated dynamical quantities such as particle number, flux, momentum,
and energy.

\subsubsection{Formalism}
We first (anti)symmetrize the elementary interaction:
\[
{\langle k_1 k_2 | \OV | k_3 k_4 \rangle} = \frac{1}{2}
(        {\langle k_1 k_2 | V | k_3 k_4 \rangle}
+ \sigma {\langle k_2 k_1 | V | k_3 k_4 \rangle})
\]
where $\sigma$ is the species flag, $-1$ for fermions and $+1$ for bosons.
From now on we deal explicitly with fermions.
Following the process outlined above in Step 2B,
the interaction piece of a Kraichnan collective Hamiltonian, in which
the generic operator from  Eq. (\ref{irr00}) is embedded, is
distinguished by the assigned set of couplings
$\varphi_{\nu_1\nu_2|\nu_3\nu_4}$ in their ${\cal M}$-fold distribution.
\begin{eqnarray}
{\cal H}^{\rm stu}_{i;{\cal N}}[\OV]
\equiv
{\cal H}_{i;{\cal N}}[\OV \varphi]
&=&
\frac{1}{2{\cal N}} \!\! {\sum_{\ell_1 \ell_2 \ell_3 \ell_4}}\!\!\!\!'
{\langle k_1 k_2 | \OV  | k_3 k_4 \rangle}~
\varphi_{\nu_1\nu_2|\nu_3\nu_4}~
\cr
&&
~~~ ~~~ ~~~ ~~~ 
\times a^{\dagger}_{\ell_1} a^{\dagger}_{\ell_2} a_{\ell_3} a_{\ell_4}.
\label{irr01}
\end{eqnarray}
These Kraichnan couplings (K couplings hereafter) may take a variety of
configurations provided their symmetry on index permutation is identical to
that of the microscopic potential $V$ in its physical indices, as in Step 1C.
Note that if $\varphi$ is an admissible K coupling then its complement
$\ophi \equiv 1 - \varphi$ is also admissible. We will develop this idea
in Sec. III in a different analysis of the exact diagrammatic expansion.

\vskip 0.15cm
Equation (\ref{irr01}) is interpreted as follows:
\begin{itemize}
\item[]
(1) The restriction $\Sigma'$ on the sum now connotes conservation of the
Kraichnan collective indices, $\nu_1 + \nu_2 = \nu_3 + \nu_4$
(modulo ${\cal N}$) as well as the physical conservation constraint on the
momenta and spins, $k_1 + k_2 = k_3 + k_4$. We have conflated index and
momentum-spin labels into the single form $\ell \equiv (k, \nu)$, so
$\ell_1 + \ell_2 = \ell_3 + \ell_4$
\cite{cons}.
\item[]
(2) The collective creation and annihilation operators $a^{\dagger}_{\ell}$
and $a_{\ell}$ are defined by Fourier sums over the distinguishable system
copies, each with its set of operators $a^{(n)\dagger}_k$ and $a^{(n)}_k$:
\begin{eqnarray}
a^{\dagger}_{\ell}
&\equiv&
\frac{1}{\sqrt{\cal N}} \sum^{\cal N}_{n=1} 
e^{2\pi i \nu n/{\cal N}} a^{(n)\dagger}_k,
\cr
a_{\ell}
&\equiv&
\frac{1}{\sqrt{\cal N}} \sum^{\cal N}_{n=1} 
e^{-2\pi i \nu n/{\cal N}} a^{(n)}_k;
~1 \leq \nu \leq {\cal N}.
\label{astra}
\end{eqnarray}
It can be shown that these collective operators satisfy the anticommutation
relations $\{a^{\dagger}_l, a_{l'}\} = \delta_{ll'}$ and
$\{a_l, a_{l'}\} = 0$.
\item[]
\vskip - 0.25cm
\hskip 0.75cm
\begin{figure}

{\includegraphics[height=5truecm]{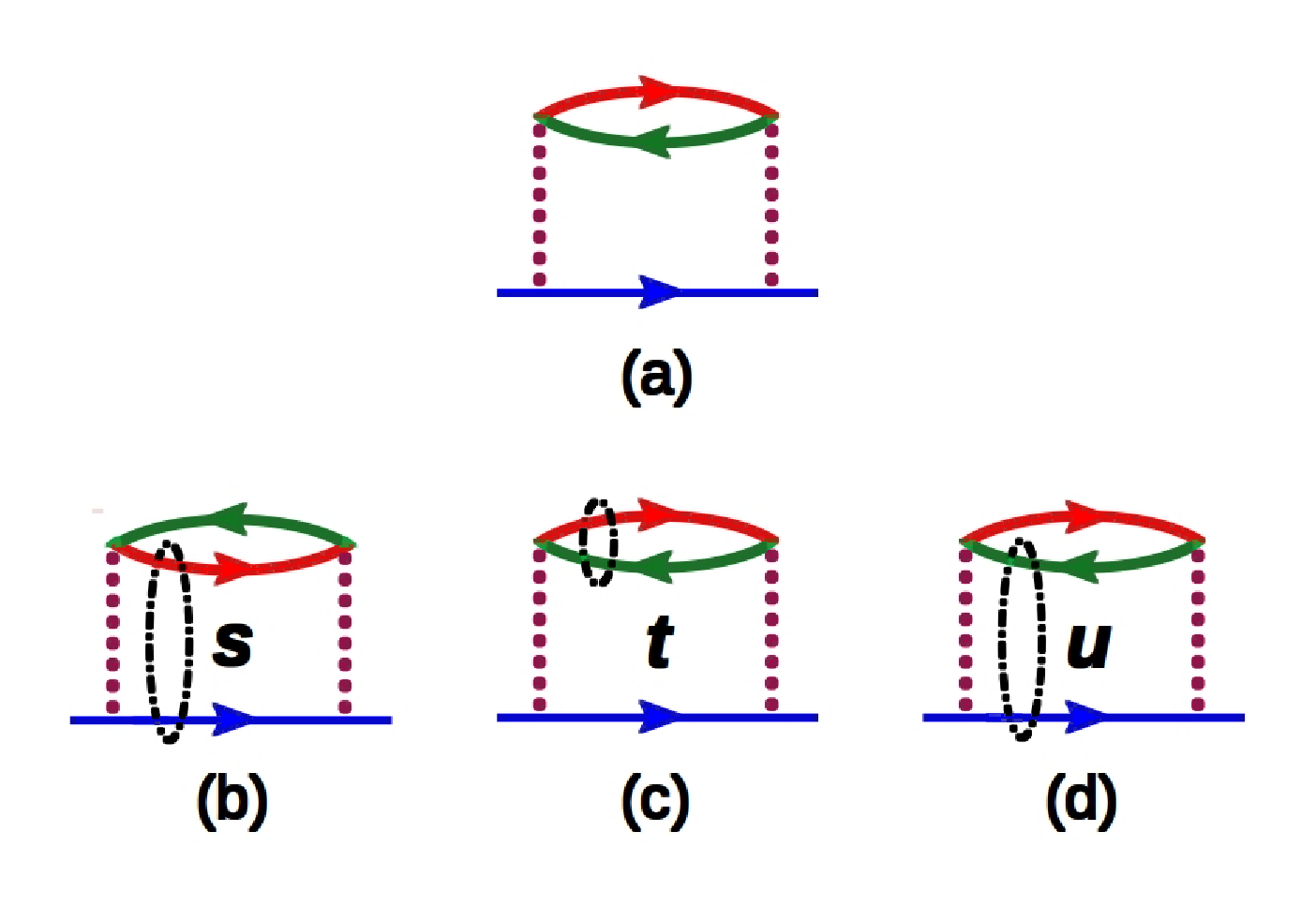}}
\caption{
Ambiguity in interpreting the second-order scattering term in the self-energy.
(a) Basic contribution; (b)--(d): each channel generates a formally diverse
but physically identical representation which, uncompensated in
Eq. (\ref{irr01}), would introduce overcounting when the second variation
$\delta^2 \Phi/\delta G \delta G$ is taken for the Kraichnan LW functional.
In the Hamiltonian ${\cal H}^{\rm stu}$ the coupling
$\varphi \equiv 1-(1\!-\!s)(1\!-\!t)(1\!-\!u)$, in place of just
$s\!+\!t\!+\!u$, inhibits redundancy while allowing free recursive interplay
of every pairing process.
The diagrams above provide
the simplest instance of structural degeneracy in a model $\Phi$ where the
nominal kernel may be equally defined to be $s-$, or $t$-, or $u$-like.
}
\label{F2}
\end{figure}

\vskip -0.15cm
(3) The Kraichnan coupling $\varphi$ comprises the $s$, $t$ and $u$ channels:
\begin{eqnarray}
s _{\nu_1 \nu_2 | \nu_3 \nu_4}
&\equiv&
\exp[i(\varsigma_{\nu_1 \nu_2} - \varsigma_{\nu_3 \nu_4})];
\cr
\varsigma_{\nu \nu'}
&\in&
[-\pi,\pi]~{\rm and}~
\varsigma_{\nu' \nu}
= \varsigma_{\nu \nu'},
\cr
\cr
t _{\nu_1 \nu_2 | \nu_3 \nu_4}
&\equiv&
\exp[i(\tau_{\nu_1 \nu_4} + \tau_{\nu_2 \nu_3})];
\cr
\tau_{\nu\nu'}
&\in&
[-\pi,\pi]~{\rm and}~
\tau_{\nu'\nu}
= -\tau_{\nu\nu'},
\cr
\cr
u _{\nu_1 \nu_2 | \nu_3 \nu_4}
&\equiv&
\exp[i(\upsilon_{\nu_1 \nu_3} + \upsilon_{\nu_2 \nu_4})];
\cr
\upsilon_{\nu \nu'}
&\in&
[-\pi,\pi]~{\rm and}~~
\upsilon_{\nu' \nu}
= -\upsilon_{\nu \nu'},
\cr
\cr
{\rm for}~~
\varphi_{\nu_1\nu_2|\nu_3\nu_4}
&\equiv&
1 - (1 \!-\! s_{\nu_1\nu_2|\nu_3\nu_4})
\cr
&&
\times
(1 \!-\! t_{\nu_1\nu_2|\nu_3\nu_4})
(1 \!-\! u_{\nu_1\nu_2|\nu_3\nu_4}).
\label{irr02}
\end{eqnarray}
This prohibits overcounting of coincident closed graphs in $\Phi$.
The need for it is shown in Fig. \ref{F2} in the context of allowing
free interplay of the three channels without duplication of
physically indistinguishable terms.
The $u$ channel has a label exchange relative to the definition of its
physical exchange counterpart, the $t$ channel. Label exchange leads to
$t_{\nu_2\nu_1|\nu_3\nu_4} \rightleftharpoons u_{\nu_1\nu_2|\nu_3\nu_4}$
while $\varphi$ always remains exchange symmetric.

\item[]
(4) A key identity in deriving the $\Phi$-derivable response within
Kraichnan's approach is
\begin{eqnarray}
\varphi_{\nu\nu'|\nu'\nu} \equiv 1
~~\text{for all}~~\nu, \nu'.
\label{irr02.1}
\end{eqnarray}
\item[]
(5) Finally, for the second embedding in the Kraichnan prescription,
Fig. \ref{F1}(b), the phase parameters $\varsigma, \tau$, and $\upsilon$
each become elements of a uniformly random distribution of size ${\cal M}$
in the limit of large ${\cal M}$.
\end{itemize}

The outcome of averaging stochastically over the distribution of $\varphi$
for a skeleton graph in the LW functional is described in Appendix B.
Channels $s, t$, and $u$ are the sole possibilities for particle-particle and
particle-antiparticle pair excitations. In terms of a system described by
pair interactions, this means that Kraichnan's construct is the most general
approximation, based on an explicit Hamiltonian, that encompasses all
possible pairwise modes.

As mentioned, the K coupling provides a new degree of freedom that lets one
preselect how the creation-annihilation operators bind to the elementary
interaction. Effectively, this fixes the possible causal orderings of the
propagator pairs at the level of the extended Hamiltonian. For conventional
parquet, the types of ordering within the pairing channels (particle-particle
or particle-hole) are specified at the level of the equations, as the
interaction $V$ itself is blind to temporal sequencing of the Green
functions. In Kraichnan's formalism the dynamical flow across the interaction
is predetermined before the equations are derived from the structure of the
extended Hamiltonian.

\subsection{Reduction of $\Phi$ by Kraichnan averaging}

\begin{figure}
\centerline{
 \includegraphics[height=12truecm]{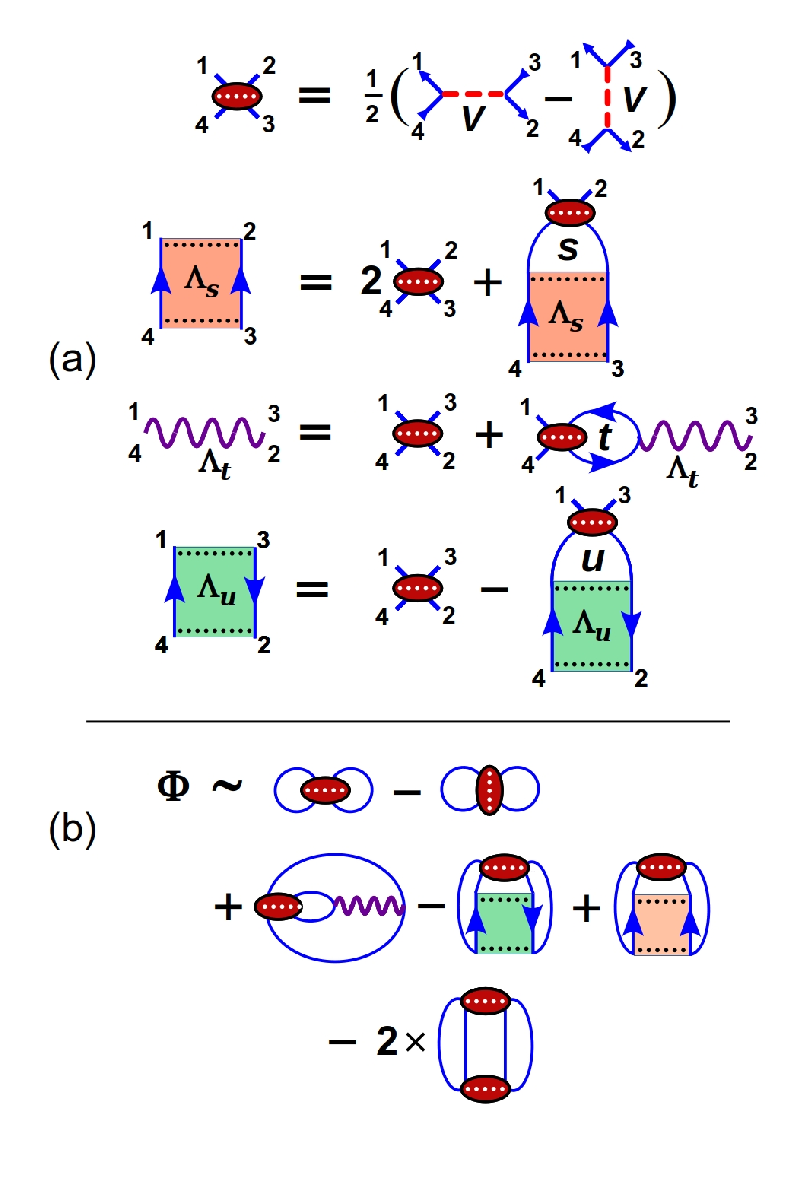}}
\caption{
(a) Definition of the primary all-order
$s, t$ and $u$ interactions. Dark ovals: antisymmetrized potential
$\OV $; linking lines are one-body propagators. Kraichnan couplings from
Eq, (\ref{irr02}), each selecting for its channel, are shown.
In the $s$ channel to leading order, the full Hartree term appears
with its Fock exchange; for $t$ and $u$ it is their superposition
as an exchange pair that generates the full Hartree-Fock term.
(b) Symbolic definition of $\Phi^{\rm stu}$, the LW correlation energy
functional (combinatorial weightings
\cite{pn}
are understood) following the Kraichnan average over all K couplings
according to Eq, (\ref{irr02}). Subtraction of two second-order skeletons
corrects for threefold overcounting in the $s,t$ and $u$ channels.
While the skeleton graphs for $\Phi^{\rm stu}$ appear simple,
their complexity lies in the selfconsistent nesting of self-energy
insertions in the one-body propagators. The $stu$ topology is fully revealed
only when the response to an external probe is extracted.
Universality of the LW functional
\cite{potthoff,lin1}
means that the topology of its constitutive diagrams in (b) is unaltered
in moving the interaction from $V$ to $\OV $ when the Hamiltonian
is itself invariant with respect to exchange. This does not affect the
overall accounting.
}
\label{F3}
\end{figure}
\vskip 0.25cm

The Luttinger-Ward functional obtained from the $stu$ Hamiltonian is 
\begin{eqnarray}
\Phi[\OV\varphi]
&=&
\frac{1}{\cal N}
\int^1_0 \frac{dz}{2z} 
{\langle \psi_0[z\OV \varphi]
\big| {\cal H}^{\rm stu}_{i;{\cal N}}[z\OV ] \big|
\psi_0[z\OV \varphi] \rangle}.
~~~ ~~~ 
\label{irr03x.0}
\end{eqnarray}
Prior to taking Kraichnan expectations, the topological content of
Eq. (\ref{irr03x.0}) remains that of the exact Eq. (\ref{aux01}), with the
addition of the collective-index degree of freedom. Since $\varphi$ has
label symmetry, $\Phi[\OV\varphi]$ is exchange invariant. Therefore so will
its Kraichnan average.

Expectations on both sides of Eq. (\ref{aux01}), over the stochastic
distribution of K couplings of Eq. (\ref{irr02}), reduce the set of LW
correlation terms to those in Fig. \ref{F3}(b) with interaction kernels defined
in \ref{F3}(a). We will denote such expectations by the subscript ${}_K$,
in which case we define
\begin{eqnarray*}
\Phi^{\rm stu}[\OV]
\equiv
{\langle \Phi[\OV\varphi] \rangle}_K.
\label{irr03y}
\end{eqnarray*}
To trace the fate of the K couplings in the upcoming variational analysis
of the $stu$ model, we consider the object $\Phi^{\rm stu}[\OV\varphi]$
although to reintroduce $\varphi$ in it is redundant.
\begin{eqnarray}
\Phi^{\rm stu}[\OV ]
&=&
\frac{1}{\cal N}
\int^1_0 \frac{dz}{2z} 
{\Big\langle{\langle \psi_0[z\OV \varphi]
\big| {\cal H}^{\rm stu}_{i;{\cal N}}[z\OV] \big|
\psi_0[z\OV \varphi] \rangle}\Big\rangle}_K
\cr
\cr
&\equiv&
\int^1_0 \frac{dz}{2z}
G[z\OV]:\Lambda[z\OV; G]:G[z\OV]
\label{irr03x}
\end{eqnarray}
where we have used the single-particle form, Eq. (\ref{irr03}).
The renormalized propagator $G$ remains defined selfconsistently by Dyson's
equation, Eq. (\ref{aux05}), now with the reduced two-body generating kernel
$\Lambda[V;G]$ of Fig. \ref{F3}(a) that includes all allowed $s,t$ and $u$
pairwise-only correlations, whose K coupling phases cancel identically to
survive averaging.

The dressed skeletons making up $\Phi^{\rm stu}$ yield a kernel $\Lambda$
that satisfies Eq. (\ref{aux05.11}) automatically since stochastic
averaging leaves that identity intact. Nevertheless, even though
$\Lambda$ is well defined and crossing symmetric, the second functional
derivative $\delta^2 \Phi^{\rm stu}/\delta G \delta G$ cannot satisfy
condition (D), Eq. (\ref{aux05.2}). Unavoidably, it generates
new structures beyond $\Lambda$. Only in the exact case (besides
Hartree-Fock, the simplest model possible) is condition (D) true
\cite{js,roger}.

The $stu$ model's violation of (D), forced by conservation, clearly
differs from parquet which does not have a counterpart to the K couplings'
structural constraints that spoil crossing symmetry. The parquet equations
do not constrain the connectivity of the channels, as $stu$ does, and
that leads to a richer family of intermediate pair scattering processes;
but parquet's greater complexity is not derived canonically from a LW
functional. Parquet is capable of inferring one by invoking Eqs.
(\ref{aux05.1}) and (\ref{aux05.11}) for its associated self-energy.

Despite the implicit appeal to a generating functional, the parquet
self-energy as such still cannot offer a way to ensure crossing symmetry
when the kernel $\delta \Sigma/\delta G$ is derived, violating condition
(D) as for $stu$. Noncrossing terms still arise, which in parquet have to
be discarded by force.

Kraichnan's Hamiltonian secures all the exact causal-analytic identities for
the reduced structures resulting from stochastic averaging. Every identity
$A = B$ depending on unitarity must hold for each collective Hamiltonian
${\cal H}^{\rm stu}[\OV \varphi]$, as for the exact one. Provided the
K coupling average is done consistently on each side of such an identity, it
follows that ${\langle A \rangle}_K = {\langle B \rangle}_K$. Relations that
depend directly on the completeness of Fock space do not survive owing to
the random-phase induced decoherence.  

\subsection{Bethe-Salpeter Equation}

The broad difference between two-body processes in $\Phi$ derivability
and parquet is that the first yields a Bethe-Salpeter equation
\cite{pn}
for the particle-hole vertex of an excitation away from equilibrium
in a strictly closed system, where particle emission cannot occur.
In parquet, the adopted extension of Bethe-Salpeter does not make a
particular distinction among particle-particle or particle-antiparticle
scattering processes. The first description has physical constraints
not necessarily applicable within the second account.

To arrive at the parquet equations' $\Phi$-derivable analogs we start by
probing the system with an external, formally nonlocal, potential
${\langle k'|U| k \rangle}$ coupling physically to each member of the
Kraichnan ensemble. It does not couple to the abstract collective indices
defined over the ensemble.

Two-body correlations unfold within their Green function like an expanding
concertina. Like a concertina, their intrinsic topology does not change
from its inner assignment in $\Phi^{\rm stu}[\OV]$. While the physical
momentum-energy flow from the outside brings out the internal dynamical
structure, the characteristic set of indexed K couplings is unchanged. What
now changes in every two-body contribution is that, in each of an infinity
of recursions, a set number of $G$ lines is singled out by the perturbation
node that each carries. This is Baym and Kadanoff's philosophy
\cite{kb1,kb2}
in Kraichnan terms.

In summary, the insertion of perturbation nodes has no effect on the
assignment of collective indices and thus on the combinatorics of
the K couplings. The physical effect is only on energy-momentum transfer
(also spin etc.).

The one-body perturbation augments the interaction
Hamiltonian, Eq. (\ref{irr01}):
\begin{eqnarray*}
{\cal H}_{i;{\cal N}}[\OV\varphi; U]
&\equiv&
\sum_{ll'} {\langle k' | U | k \rangle} a^{\dagger}_{l'} a_l
+ {\cal H}_{i;{\cal N}}[\OV\varphi].
\end{eqnarray*}
Response to a local field is generated by setting
${\langle k' | U | k \rangle} \to U(q,\omega)\delta_{k',k+q}$,
dynamically linking (contracting) the particle-hole propagators
that terminate and start at $U$.

\begin{figure}
\centerline{
 \includegraphics[height=2.25truecm]{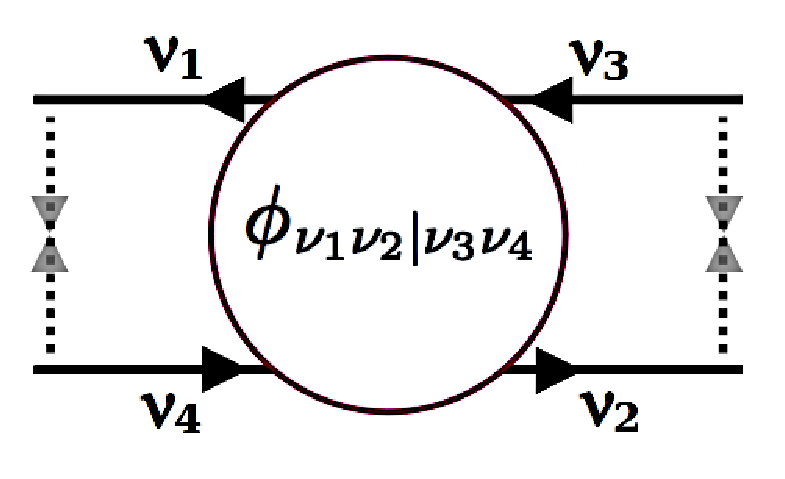}}
\caption{
Index convention for a Kraichnan vertex, associating
with it the nominal K coupling $\phi_{\nu_1\nu_2|\nu_3\nu_4}$. A response
term results when $G$ lines are attached left and right and join at a
perturbation node. The diagram contributes if and only if the internal
sum of coupling phases cancels on connection to the effective
K coupling $\phi_{\nu_1\nu_2|\nu_3\nu_4}\rightarrow
\varphi_{\nu_1\nu_2|\nu_2\nu_1} \equiv 1$ as in Eq. (\ref{irr02.1}).
This construct manifests the same closed topology previously
implicit in the unitary structure of $\Phi$.
}
\label{F4}
\end{figure}

Next we take physical expectations only, retaining the explicit K couplings
to keep track of all potential pair processes before Kraichnan averaging.
We sketch the logic
\cite{KII}.
The two-body Green function is $\delta G/\delta U$
\cite{kb1}; 
note, quite generally, that $U$ adds linearly to $\Sigma$ so
$\delta \Phi/\delta U = \delta \Phi/\delta \Sigma = G$ so
$\delta G/\delta U = \delta^2 \Phi/\delta U \delta U$.

Varying both sides of the Dyson equation (\ref{irr03}) produces
\begin{eqnarray}
\frac{\delta G(21)}{\delta U(56)}
&\equiv&
G(25)G(61)
+ G(21')G(2'1)
\cr
&&
\times
\varphi_{\nu_{1'}\nu_3|\nu_{2'}\nu_4}\Lambda'(1'3|2'4)
\frac{\delta G(43)}{\delta U(56)}
\label{kII18.2}
\end{eqnarray}
where we define $\Lambda'$ via $\varphi\Lambda'\equiv\delta\Sigma/\delta G$,
being  accompanied by an effective K coupling as for Fig. \ref{F4}.
The functional equation remains completely general, applicable to any
suitable choice of Kraichnan coupling including the exact case
$\varphi \equiv 1$.

In shorthand, with $I$ the two-point identity matrix,
the Neumann series for $\delta G/\delta U$ becomes
\begin{eqnarray}
\frac{\delta G}{\delta U}
&=&
[II - GG\!:\!\varphi\Lambda']^{-1}\!:\!GG
\cr
&=&
GG + GG\!:\!\varphi\Lambda'\!:\![II - GG\!:\!\varphi\Lambda']^{-1}\!:\!GG.
\label{kII18.3}
\end{eqnarray}
Recalling Fig. \ref{F3}(a), the form of the generating kernel $\Lambda$
for purely $stu$ correlations, namely for the reduced LW functional
$\Phi^{\rm stu}[\OV\varphi]$, can be read off in terms of the subsidiary
kernels $\Lambda_s, \Lambda_t$ and $\Lambda_u$ for each channel. To put all
interactions on the same K coupling-free footing as $\OV$ we write $\phi$
for the latter's coupling and factor it out; refer also to Fig. \ref{F4}.
Any surviving chain of K couplings, whose phases cancel right across, finally
merges constructively with $\phi^{-1}$ as exemplified in Fig. \ref{F3}(b).
In the response description the unpaired outermost indices will be
contracted when the terminating lines $G$ link to the perturbation
nodes, since only for equal indices across a node is there a  nontrivial
Kraichnan expectation. (Figures \ref{F6} and \ref{F7} below
hold more details.) Then
\begin{eqnarray}
\Lambda
&=&
2\OV 
+ {\phi}^{-1} {\left(
  \OV\varphi :GG: s\Lambda_s
+ \OV\varphi :GG: t\Lambda_t
\right.}
~~~ ~~~ 
\cr
&&
{\left.
-~ \OV\varphi :GG: u\Lambda_u
\right)}
\cr
{\rm where}~
\Lambda_s
&\equiv&
2\OV + {\phi}^{-1}\OV\varphi :GG: s\Lambda_s,
\cr
\Lambda_t
&\equiv&
\OV  + {\phi}^{-1}\OV\varphi :GG: t\Lambda_t
\cr
{\rm and}~
\Lambda_u
&\equiv&
\OV  - {\phi}^{-1}\OV\varphi :GG: u\Lambda_u.
\label{kII15}
\end{eqnarray}
The different accounting for $\OV$, which applies likewise to the
kernel equations to follow, is because the $s$ channel incorporates its own
ladder exchange while $t$ and $u$ are each other's distinct exchanges;
only if summed would they carry the full Hartree-Fock interaction $2\OV$.
The last three relations in Eq. (\ref{kII15}) express the content
of Fig. \ref{F3}(a).
\begin{figure}
\centerline{
 \includegraphics[height=7.5truecm]{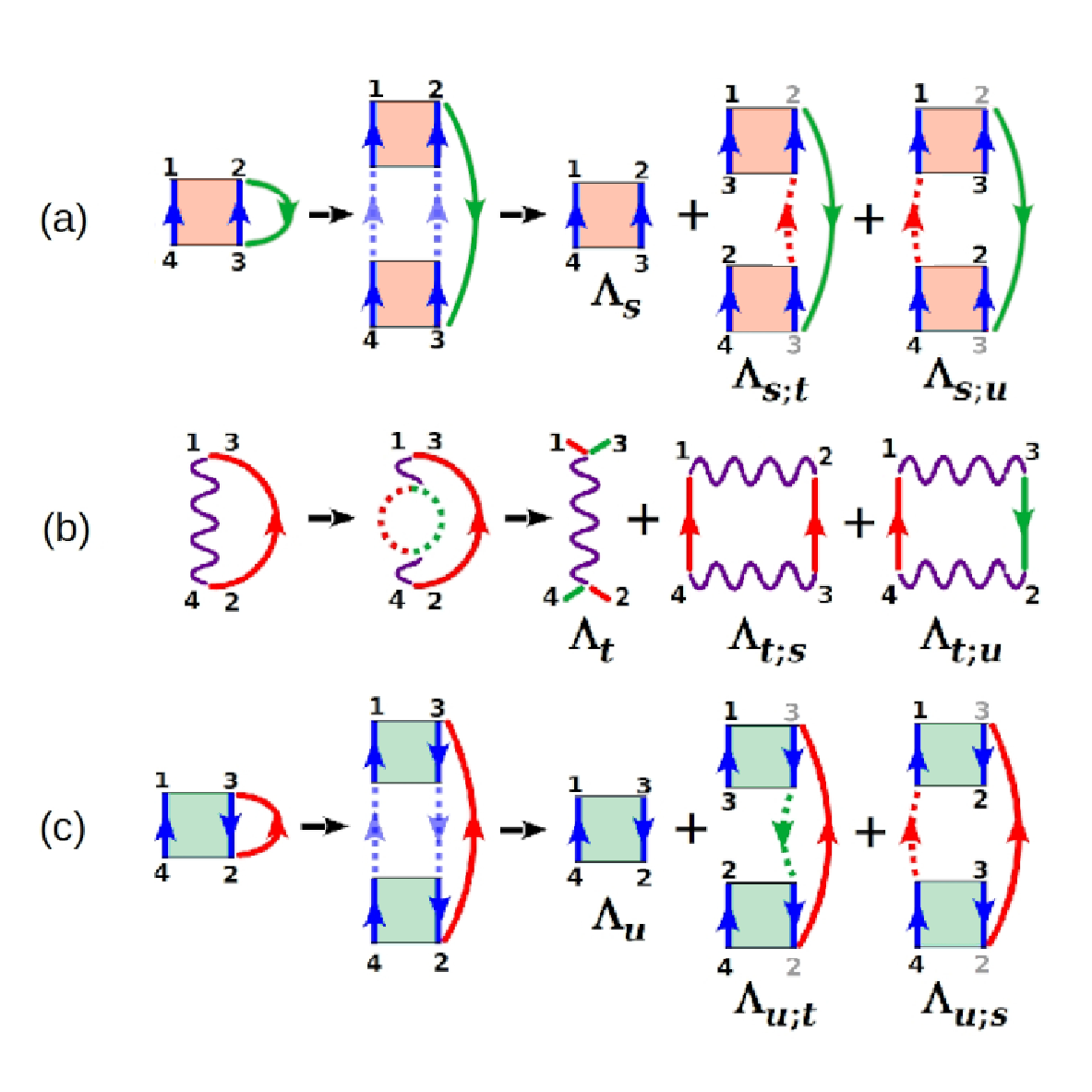}}
\caption{
Systematic removal of a propagator $G$ internal to the self-energy
$\Sigma[\varphi\overline V; G] = \Lambda\!:\!G$ after Kadanoff and Baym
\cite{kb1,kb2}. 
This generates the primary $stu$ scattering kernel
$\Lambda' = \delta^2 \Phi^{\rm stu}/\delta G \delta G$.
Removing $G(32)$, solid line,
simply regenerates $\Lambda$. Removing any internal $G$ lines (dotted) other
than $G(32)$ yields the additional vertices required by microscopic
conservation. (a) Beyond the $s$-channel ladder $\Lambda_s$ the noncrossing
symmetric $t$-like term $\Lambda_{s;t}$ and $u$-term $\Lambda_{s;u}$ are
generated.
(b) Generation of $\Lambda_t$ and the nonsymmetric $\Lambda_{t;s}$,
$\Lambda_{t;u}$. (c) Generation of $\Lambda_u$ with $\Lambda_{u;t}$ and
$\Lambda_{u;s}$. No diagrammatic structure emerges that is not already
incorporated recursively in the propagators $G$ within $\Phi$.
Note that $\Lambda' \!-\! \Lambda$ starts at {\em third order} in $\OV$.
}
\label{F5}
\end{figure}
\vskip 0.15cm

\subsection{Conservation in $\Phi$ derivability, and its cost}

We face the inevitable outcome of every $\Phi$-derivable model: unlike
Item (D) for the exact theory, the response kernel $\delta \Sigma/\delta G$
carries new terms beyond $\Lambda$. The process is shown graphically in
Fig. \ref{F5}. These apparently extraneous vertices are essential to consistency
in conservation for the two-body Green function $\delta G/\delta U$, but
are no longer crossing symmetric and therefore cannot contribute to the
parquet scattering amplitude as conventionally understood
\cite{pqt3,roger}. 

From Eq. (\ref{kII18.3}) the complete four-point kernel is defined:
\begin{eqnarray}
\Gamma'
&\equiv&
{\phi}^{-1}\Lambda'\varphi\!:\![II - GG\!:\!\varphi\Lambda']^{-1}.
\label{kII18.4}
\end{eqnarray}
Consequently the conserving two-body Green function is
\begin{eqnarray}
\frac{\delta G}{\delta U}
&=&
GG\!:\![II + \varphi\Gamma'\!:\!GG].
\label{kII18.5}
\end{eqnarray}
Momentum transfer in the above is determined by all the
K couplings attached to each interaction, not by imposing the proper
analytic form on intermediate pairs $GG$. This differs from the
expansion of the standard parquet vertex
\cite{pqt3},
in which the mode of transfer has to be specified explicitly for the
each of the three possible pairs $GG$.

The task is to show the equivalence between the system of Equations
(\ref{kII18.3}) to (\ref{kII18.5}) on the one hand and, on the other, the
coupled parquet-like equations emerging from the Kraichnan formalism.

\subsection{$\Phi$ derivability in the Kraichnan representation}

The K coupling constraints make a radical change to the physical
content of the parquet equations' $\Phi$-derivable form, in contrast with
their conventional presentation. Before deriving them we clarify the
one-to-one correspondence between the Baym-Kadanoff and Kraichnan
interpretations of response.

\begin{figure}
\centerline{
 \includegraphics[height=6truecm]{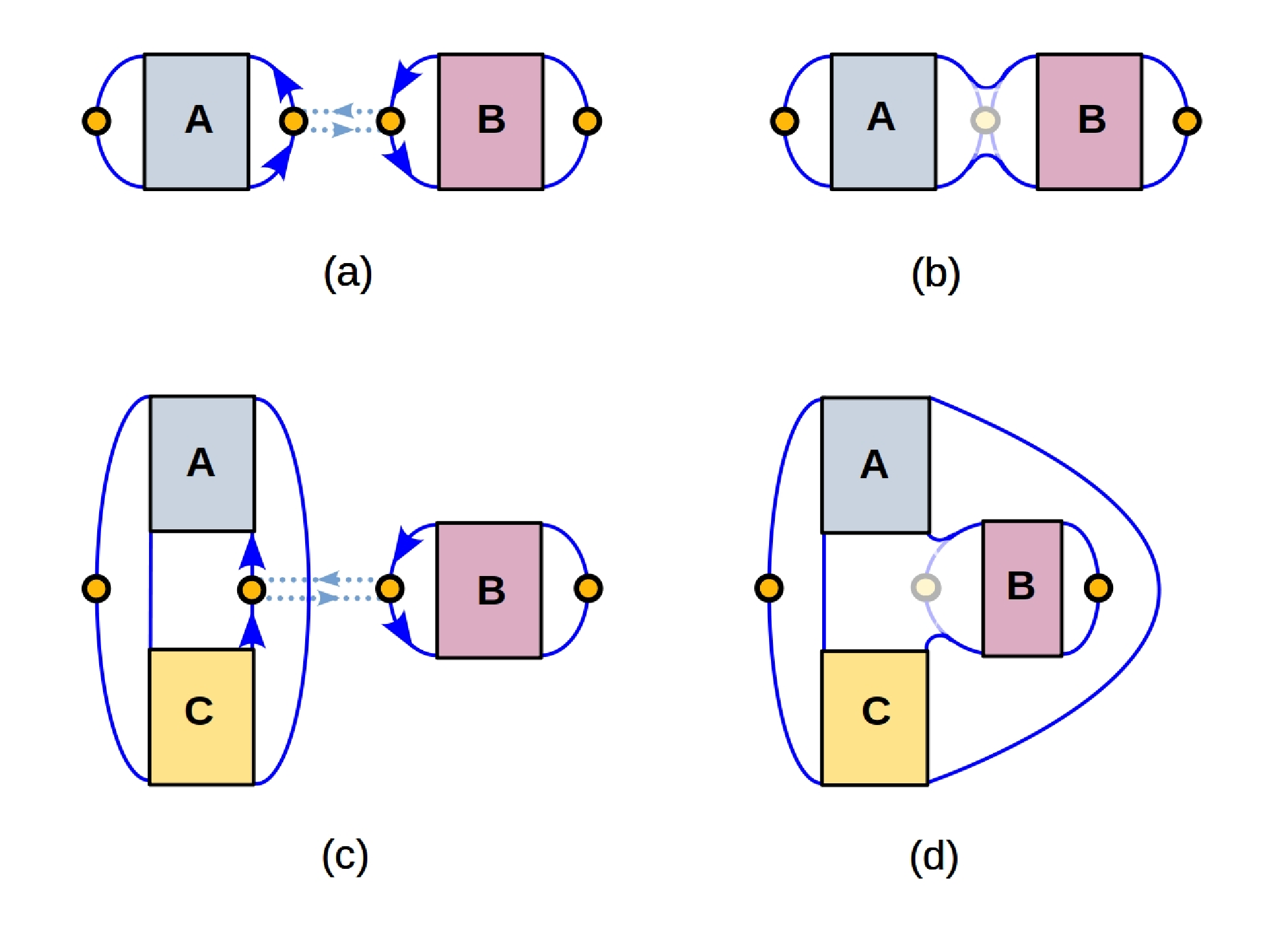}}
\caption{
Recursive construction of response kernel within
$\Phi$ derivability. Dots: external perturbation nodes. (a) Two
contributions A and B to the particle-hole response function combine into
(b), a new contribution. Graphs (c) and (d) show a more complex combination
with a third allowed contribution C.
In the response description Eq. (\ref{kII18.2}), the system cannot tell
a direct perturbation from one that is induced, so a perturbation node may be
freely replaced with an induced perturbation. Fusion of the response terms
produces a new contribution to the total. In the process the internal
topology of the resultant response, virtual within the renormalization of
$\Phi$, becomes manifest. All the kernel parts in Fig. \ref{F5} are
recursively convolved in this way.
}
\label{F6}
\end{figure}
\vskip 0.15cm

In Fig. \ref{F6} we illustrate how the standard Baym-Kadanoff derivation
implements the total system response as the sum of the direct response to
the external perturbation and all the nonequilibrium fluctuations induced
by it within the correlated system. Since the component particles cannot
distinguish between direct and induced disturbances, the response to every
such dynamical stimulus is highly selfconsistent. It follows that the
primitive components of the response become convolved,
leading to the systematic cascade of contributions implemented in
Eq. (\ref{kII18.2}).

\begin{figure}
\centerline{
 \includegraphics[height=6truecm]{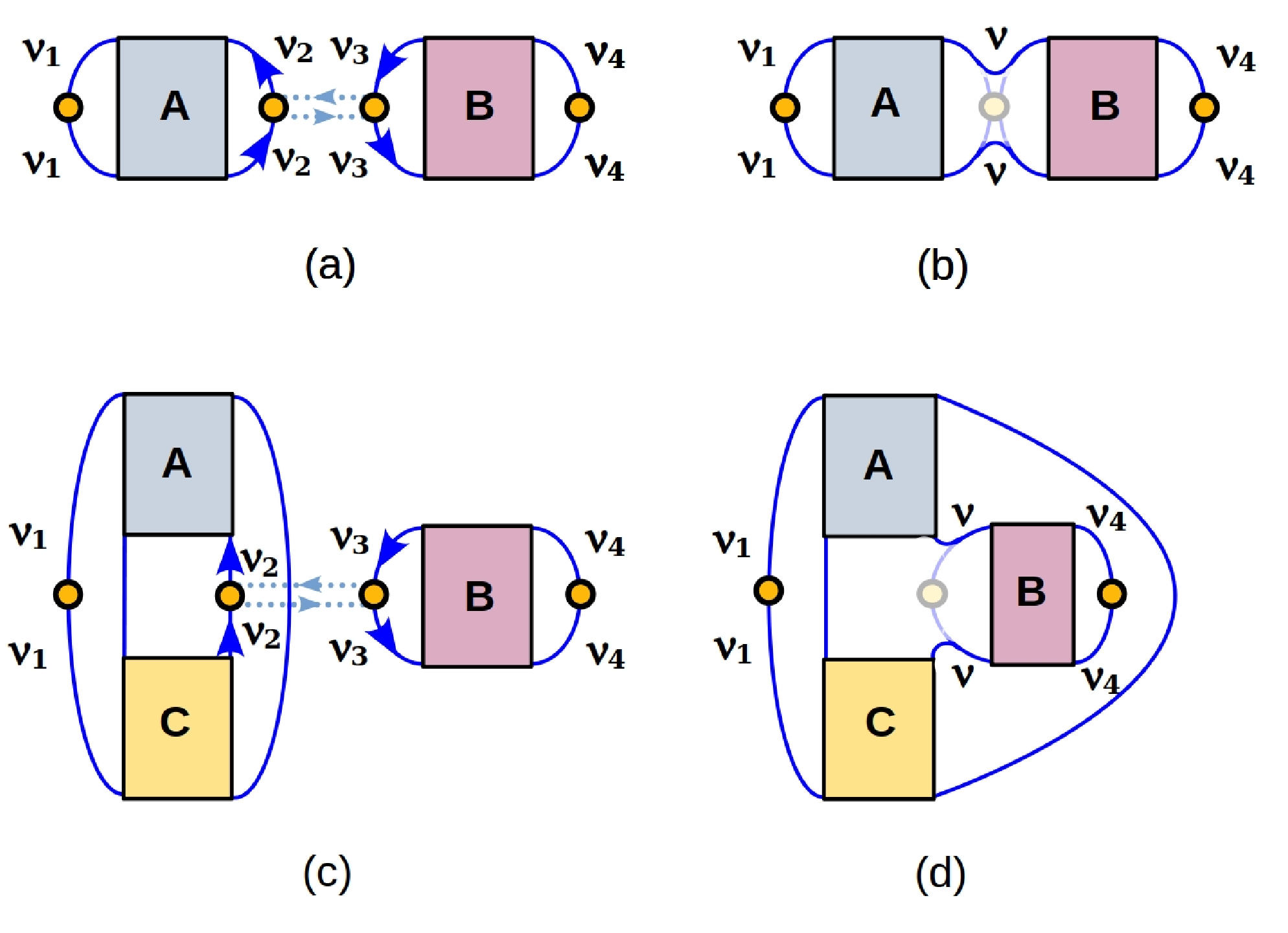}}
\caption{
Recursion of the response kernel in the Kraichnan
approach, functionally equivalent to Fig. \ref{F6}. Terms (a) and (c) go to (b) and
(d) respectively; their fusion is now dictated by global index-sum
conservation. In (a), terms A and B survive Kraichnan averaging separately.
Since their internal phases are unaffected by their fusion to form (b), the
new term trivially survives averaging; likewise the case of (c) going to (d).
The outcome maps exactly onto the $\Phi$-derivable structure of Fig. \ref{F6},
replicating the topology of all contributions to the response
$\delta^2 \Phi/\delta U \delta U$.
}
\label{F7}
\end{figure}

Construction of the response within Kraichnan's canonical formulation is
expressed differently but describes the very same processes as the more
heuristic $\Phi$-derivable description.
The rules, analogous to physical conservation, are (\i) conservation of
incoming and outgoing index sums across any pair-scattering amplitude
(alongside conservation of momentum) and (\i\i) conservation of index across
any perturbation node. Figure \ref{F7} replicates Fig. \ref{F6} in these
terms. An alternative criterion for a candidate response term is whether it
can be recollapsed to a diagram of $\Phi$.

\vskip 0.15cm

A different way to state the above is that the phase structure of a
Kraichnan-embedded diagram within the LW functional, not just skeletal but
recursively defined, cannot change under an external stimulus. While the
input of external energy and momentum reveals the distinctly dynamical
correlations (otherwise implicit at equilibrium), only those correlations
with the same K coupling topology, showing overall phase cancellation, will
survive as response contributions in the form of Figs. \ref{F6} and \ref{F7}.

With respect to Fig. \ref{F7} in particular it is noteworthy that, for a
convolution of two independently closed response diagrams within a composite
term, the overall phase cancellation for the interactions of one component
is completely disjoint from the corresponding configuration in the other
component. Indeed, the disjoint parts could each have a completely different
coupling scheme without affecting the accounting of its complement. This
separability will be exploited in the following Section.
Figure \ref{F7} shows how phase-sum decoupling is induced by the
conservation rules for the collective indices.

\subsection{Minimal parquet: $stu$ version}

Within the $stu$ approximation, Eq. (\ref{kII18.5}) contains all conserving
dynamical processes involving two-body propagation between interactions. Its
form, however, obscures the distinct action of the three channels $s,t$ and
$u$, as evidenced in classic parquet. To recast Eq. (\ref{kII18.5}) and reveal
its parquet-like architecture, we review the derivation by K coupling analysis.

To the bare potential $\OV $ as the primary irreducible starting term, the
extra noncrossing symmetric term $\Lambda'' \equiv \Lambda' - \Lambda$ has to
be added. This could not appear in normal parquet as it violates crossing
symmetry for fermions. Were it crossing symmetric, then $\Lambda'$ itself
would be crossing symmetric as well as conserving and therefore exact
\cite{roger};
but precisely because $stu$ is conserving, $\Lambda''$ is needed to construct
the appropriate two-body Green function and system response. Let
\begin{eqnarray}
{\cal V} \equiv \OV  + \Lambda''.
\label{kII15.4}
\end{eqnarray}
Three auxiliary kernels are defined. Each is irreducible in the channel for
which it is the input. Each carries only terms that do not propagate
directly in the given channel. That is, the kernel's terms can be split
only by cutting $GG$ pairs propagating in the other two channels.

The auxiliary kernels drive the
equation for the complete vertex $\Gamma'$, fed into the two-body Green
function in Eq. (\ref{kII18.5}). When contracting this to obtain the
response, stochastic averaging of product sequences of K couplings keeps
those terms compatible with unitarity.
As part of the seed kernel, $\OV$ will regenerate $\Lambda$ itself while
$\Lambda''$, whose expansion starts at third order in $\OV$, generates all
the additional response terms need for conservation.
The $s$-, $t$- and $u$-irreducible auxiliary kernels are
\begin{eqnarray}
\Gamma'_s
&\equiv&
\OV + {\cal V}
+ {\phi}^{-1} (
  \Gamma'\varphi :GG: t\Gamma'_t
- \Gamma'\varphi :GG: u\Gamma'_u
);
~~~ 
\cr
\Gamma'_t
&\equiv&
{\cal V}
+ {\phi}^{-1} (
- \Gamma'\varphi :GG: u\Gamma'_u
+ \Gamma'\varphi :GG: s\Gamma'_s
);
\cr
\Gamma'_u
&\equiv&
{\cal V}
+ {\phi}^{-1} (
  \Gamma'\varphi :GG: s\Gamma'_s
+ \Gamma'\varphi :GG: t\Gamma'_t
)
\label{kII15.5}
\end{eqnarray}
which combine to yield the complete response kernel
\cite{xsym}:
\begin{eqnarray}
\Gamma'
&=&
\OV + {\cal V}
+ {\phi}^{-1} (
  \Gamma'\varphi :GG: s\Gamma'_s
+ \Gamma'\varphi :GG: t\Gamma'_t
~~~ ~~~ 
\cr
&&
- \Gamma'\varphi :GG: u\Gamma'_u
),
\label{kII15.6}
\end{eqnarray}

Beyond partitioning the response kernel in terms of its (mutually)
irreducible pairwise processes, there is a subtle procedural distinction
between the Kraichnan perspective of Eqs. (\ref{kII15.5}) and (\ref{kII15.6})
and the Kadanoff-Baym one of Eqs. (\ref{kII18.4}) and (\ref{kII18.5}). The
latter starts from an already approximate form for the LW functional and
proceeds by tracking its selfconsistently recursive structure, Fig. \ref{F6}.
In the former approach everything stays exact prior to stochastic averaging
whereas, in our derivation of the parquet equations, we have quietly neglected
every contribution that is not pairwise linked in the pre-average collective
Kraichnan LW functional. The {\em a posteriori} justification is that, in
any case, stochastic averaging projects out only the purely pairwise $stu$
correlations.

In $\Phi$-derivable models $\Gamma'$ is not the main goal. The closure
procedure in Fig. \ref{F7}, tied to the perturbation nodes, is regulated by
the outermost K coupling $\phi$. The presence of this overarching constraint
says that the open links in the vertices of Eq. (\ref{kII15.6}) have a
different role in $\Phi$ derivability from their unconstrained analogs in
standard parquet, with its S-matrix view.

Crossing symmetry plays an indirect role in response, shaping the four-point
kernel in the abstract. There is consistency with Pauli exclusion but no
direct connection to the response obtained from the general two-body Green
function. Probing a closed system involves directly exciting particle-hole
pairs rather than particle-particle. A notable example of a noncrossing
symmetric yet physically justified $\Phi$-derivable model is the random-phase
approximation
\cite{dp},
with exchange having no part in an essentially long-wavelength description.

There is a significant point to make about the compensating function of the
K couplings on the right-hand side of Eq. (\ref{kII15.6}). Standard parquet
has no counterpart to $\varphi$. As Fig. \ref{F2} shows, however, in any
reconstruction of the LW functional starting from a kernel, the three
different pairing channels result in the same contribution to $\Sigma$ and
thus $\Phi$. The K coupling within $\Phi$ is defined to prevent such
overcounting. Therefore recovering a model LW functional, by working up
from the parquet equations, means introducing by hand a subtraction that
would remove the threefold redundancy in $\Phi$.

\section{Irreducibility and Exact Parquet}

Kraichnan's formalism provides a systematic procedure, a stochastic
algorithm, to isolate every irreducible contribution to the exact
Luttinger-Ward functional that has no description in purely pairwise $stu$
terms. These will convolve naturally with the strictly $stu$ correlations
to recover the canonical LW description and lead to a different formulation
of the exact parquet equations.

Recall that, prior to taking K coupling expectations, the extended Kraichnan
Hamiltonian retains the functional structure of its original. The only
distinction between the pair-only Luttinger-Ward $\Phi^{\rm stu}$ and the
$stu$ irreducible complement, call it $\Phi^{\rm cmp}$, lies in how the
latter's K couplings are defined.

The correlation structure beyond $stu$ must be governed
by a Kraichnan coupling complementary to $\varphi$, or
\begin{eqnarray}
\ophi
&=&
1 - \varphi \equiv \os \ot \ou
\cr
{\rm where}~~
\os
&\equiv&
1\!-\!s,~ \ot \equiv 1\!-\!t ~{\rm and}~\ou \equiv 1\!-\!u.
\label{bars}
\end{eqnarray}
Contributions to the corresponding LW functional $\Phi^{\rm cmp}$ include all
the $stu$-irreducible terms to all orders in $\OV$ beyond the leading
Hartree-Fock diagrams. With no loss of generality $\ophi$ can be recast.
To a K coupling $\varphi_{\nu_1\nu_2|\nu_3\nu_4}$ associate the K coupling
$\varphi' \equiv \varphi_{\nu_1\nu_2|\nu_2\nu_1}$, which from
Eq. (\ref{irr02.1}) is identically unity. Writing
$\ophi = \varphi' - \varphi$ lets one conveniently treat both
K couplings and ``anticouplings'' uniformly. Manifestly, the indices
of $\varphi'$ will pass unchanged across the associated interaction,
as if the latter were absent from the Kraichnan point of view.

In constructing $\Phi^{\rm cmp}$ we note that its kernel, call it $\Xi$,
plays the role of $\Lambda$ in $\Phi^{\rm stu}$ but now selects all possible
non-pairwise-connected skeletons from the exact LW functional; since
$\Phi^{\rm cmp}$ retains exchange invariance, a crossing symmetric
$\Xi$ exists.

The complementary Hamiltonian is
\begin{eqnarray*}
{\cal H}^{\rm cmp}_{i;{\cal N}}[\OV ]
&\equiv&
{\cal H}_{i;{\cal N}}[\OV \ophi]
\cr
&=&
\frac{1}{2{\cal N}} {\sum_{\ell_1 \ell_2 \ell_3 \ell_4}}\!\!\!'
{\langle k_1 k_2 | \OV  | k_3 k_4 \rangle}~
a^{\dagger}_{\ell_1} a^{\dagger}_{\ell_2} a_{\ell_3} a_{\ell_4}.
\cr
&&
~~~ ~~~ ~~~ \times
\os_{\nu_1\nu_2|\nu_3\nu_4} \ot_{\nu_1\nu_2|\nu_3\nu_4} \ou_{\nu_1\nu_2|\nu_3\nu_4}.
\end{eqnarray*}
On K-averaging this leads to the definition of the
complementary LW functional 
\begin{eqnarray}
\Phi^{\rm cmp}[\OV]
&\equiv&  
\int^1_0 \frac{dz}{2z}
{\langle G:{\cal H}^{\rm cmp}_{i;{\cal N}}[z\OV]:G \rangle}_K
\cr
&=&
\int^1_0 \frac{dz}{2z}
G:\Xi[z\OV;G]:G
\label{irrX1.0}
\end{eqnarray}
where, as in Eq. (\ref{irr03x}), the Kraichnan-averaged propagator is
renormalized in keeping with the reduced diagrammatics of this particular
model. The counterparts to Eqs. (\ref{kII18.4}) and (\ref{kII18.5}) apply,
with the residual kernel $\Xi''=\delta^2\Phi^{\rm cmp}/\delta G\delta G-\Xi$
no longer crossing symmetric, as for $\Lambda''$ earlier.

\subsection{Interplay of $stu$ and complementary sectors}

\begin{figure}
\centerline{\hskip 8mm 
 \includegraphics[height=7truecm]{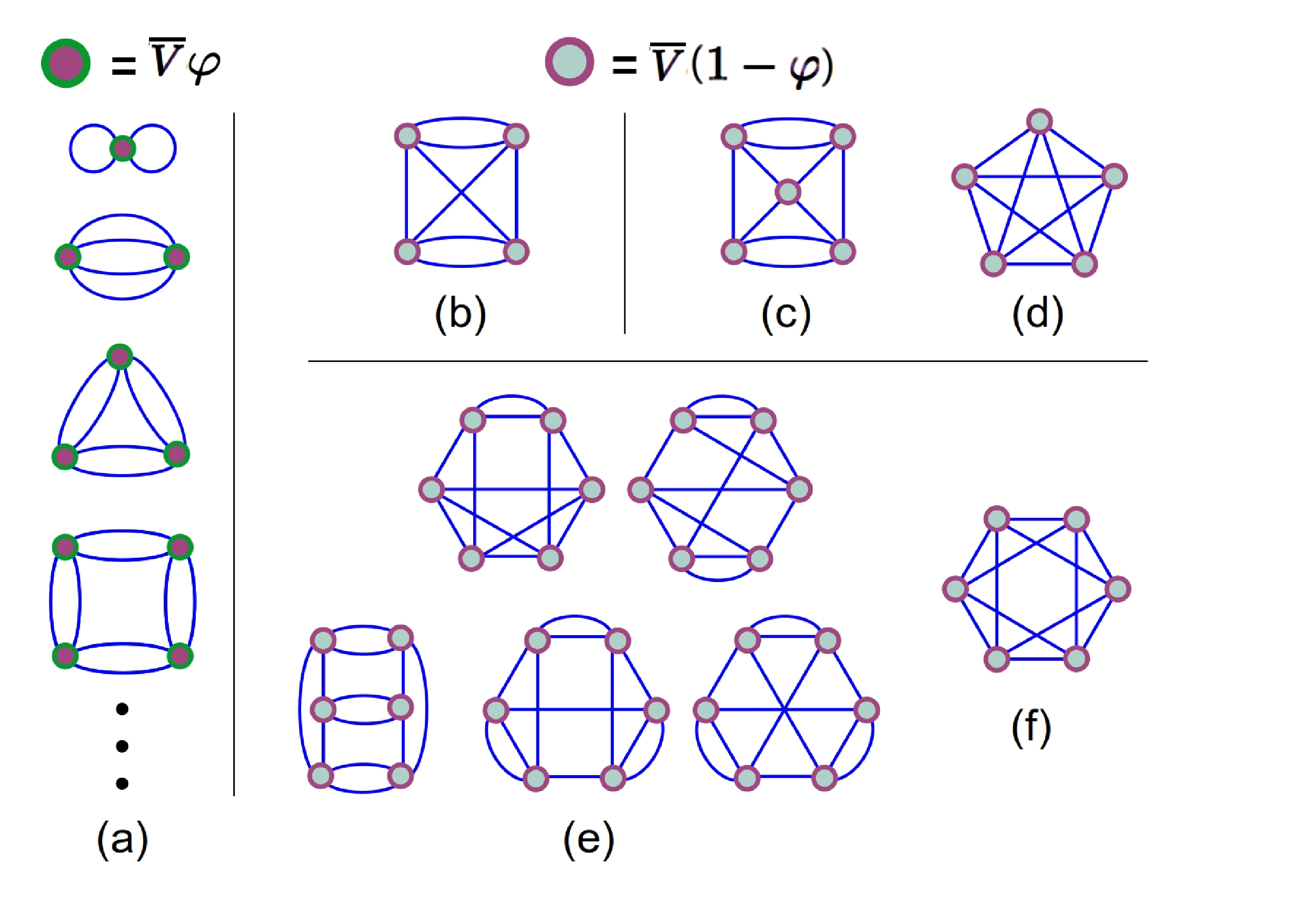}}
\caption{
Comparison of $stu$ reducible and irreducible skeletons in the expansion of
the Luttinger-Ward functional. (a) Closed $stu$ diagrams comprise those of
Fig. \ref{F3}(b): every node, dark dots, is a K-coupled interaction
$\OV\varphi$. (b) Non-$stu$ term at fourth order in the complementary
node $\OV(1-\varphi)$ as light dots. (c) Fifth-order non-$stu$ skeleton,
(d) fifth-order primitive irreducible that, unlike (b), (c) and (e)
cannot be split by cutting two line pairs (other than by trivial removal
of a node). (e) Non-$stu$ two-pair reducibles at sixth order;
(f) primitively irreducible sixth-order graph.
}
\label{F8}
\end{figure}
\vskip 0.15cm

In Fig. \ref{F8} we show representative diagrams of the full
Luttinger-Ward functional, both $stu$-reducible and -irreducible.
Here the three channels $s, t$ and $u$ from Fig. \ref{F3} are conflated
and characterized by the single node $\OV\varphi$ which freely selects
all possible modes for exclusively pairwise transfer of conserved quantities.
The $stu$-irreducibles, by definition, cannot be represented in this way;
their extended coupling is set by the Kraichnan-modified interaction
$\OV\ophi$.

Unlike Figs. \ref{F8}(d) and (f), not all non-$stu$ diagrams appear to
fulfill the requirement (C) for symmetry, as met both by the exact
four-point kernel and the reduced $stu$ series, recast as Fig. \ref{F8}(a).
The lack of full symmetry among propagators is seen in Figs. \ref{F8}(b),
(c) and (e). Resolution of the apparent violation of requirement (C) for some
non-$stu$ graphs is found in Appendix A, which discusses the appropriate
numerical weighting in evaluating $\delta \Phi/\delta G \equiv \Sigma$.
The claim in Ref. \cite{KII} that all closed diagrams should display perfect
equivalence of its single-particle lines, is too restrictive. 

From Eq. (\ref{bars}) it follows that  $\Phi^{\rm stu}$ and $\Phi^{\rm cmp}$
are mutually exclusive. Take any $stu$ skeleton of order $n$ chosen from
those in Fig. \ref{F8}(a) but with $\OV\ophi$ replacing each $\OV\varphi$.
The product of complementary K couplings becomes
\begin{eqnarray}
\prod^n_{i=1}(\varphi'_i - \varphi_i)
&=&
  (-1)^n\prod^n_{i=1}\varphi_i
+ (-1)^{n-1}\sum^n_{j=1}\varphi'_j\prod_{i\neq j}\varphi_i
~~~ ~~~ 
\cr
+~ (-1)^{n-2}
&&
\!\!\!\!\!\! \!\!\!\!\!\! \!\!\!\!
\sum^n_{j<k} \varphi'_j\varphi'_k
\prod_{j\neq i\neq k}\!\!\! \varphi_i
+ ...
+ \prod^n_{i=1}\varphi'_i.
\label{xi01}
\end{eqnarray}
The Kraichnan average is taken for each product of factors
on the right-hand side of Eq. (\ref{xi01}). At every occurrence of
$\varphi'_i$, the indices do not change in crossing the interaction and the
total number of free indices decreases by two. Therefore, as every freely
indexed factor $\varphi_i$ in Eq. (\ref{xi01}) is replaced with its
constrained partner $\varphi'_i$, the same order-$n$ skeleton graph
is in play but is averaged only for the analogous order-$m$ product
of free $\varphi_i$ while carrying the sign factor $(-1)^m$,
where $0\leq m\leq n$.

Combinatorially, the total of the identical $stu$ skeletons contributing
to the K-average over the assembly of terms on the right-hand side
of Eq. (\ref{xi01}) is
\begin{eqnarray*}
\sum^n_{m=0}
{\left(
\begin{matrix}
n \cr
m
\end{matrix}
\right)}
(-1)^m
&=&
(1 - 1)^n = 0.
\label{xi02}
\end{eqnarray*}
Hence $\Phi^{\rm cmp}[\OV\ophi]$ excludes all $stu$ skeletons. The result
holds in the presence of an external perturbation $U$, namely for terms as
in Fig. \ref{F7}, so all diagrams for the two-body Green function
$\delta^2\Phi^{\rm stu}[\OV\varphi]/\delta U\delta U$ are also excluded
from its counterpart $\delta^2\Phi^{\rm cmp}[\OV\ophi]/\delta U\delta U$.
The exclusion is mutual, since a complementary element could not be in the
$stu$ set without contradiction by the foregoing argument.

\subsection{Exact expansion in the Kraichnan representation}

The expectations over both types of K coupling, each allowed to act
in its own right in the context of the structures of
${\cal H}^{\rm cmp}_{i;{\cal N}}[\OV]$ and
${\cal H}^{\rm stu}_{i;{\cal N}}[\OV]$, recover the exact form of $\Phi$
because every closed nonpairwise graph suppressed by averaging in the
original $stu$ mode, is restored on averaging in the complementary mode
but crucially now in the full presence of the pairing dynamics embodied
in the generic Eq. (\ref{kII18.4}).
Write the exact Hamiltonian as
\begin{eqnarray}
{\cal H}_{i;{\cal N}}[\OV]
&\equiv&
{\cal H}_{i;{\cal N}}[\OV (\ophi \!+\! \varphi)]
=
  {\cal H}^{\rm cmp}_{i;{\cal N}}[\OV]
+ {\cal H}^{\rm stu}_{i;{\cal N}}[\OV];
~~~ ~~~ 
\label{irrX1.1}
\end{eqnarray}
the exact LW functional will be independent of $\varphi$.

In the Kraichnan-based analysis of the exact correlation functional we
retain the explicit structure of $\Phi^{\rm cmp} \sim G:\Xi:G$ from
Eq. (\ref{irrX1.0}) to act as the seed kernel for the expansion in terms
of the pair channels. We then define the one-pair reducible complement
\[\Upsilon[\OV;G] \equiv \Gamma[\OV;G] - \Xi[\OV;G].
\]
Since ${\cal H}^{\rm stu}$ acts as a strong perturbation on
${\cal H}^{\rm cmp}$, new hybrid terms are generated so the perturbative
expansion of $\Upsilon$ itself, while consisting of pairwise connected
elements, will no longer be just $\Lambda$ from Eq. (\ref{irr03x})
and Fig. \ref{F3}. It will involve autonomous $stu$ components convolved
with autonomous non-$stu$ ones, coupled via the same index conservation
principle illustrated in Fig. \ref{F7}.

The exact Luttinger-Ward functional takes on the form
\begin{eqnarray}
\Phi[\OV]
&=&
\frac{1}{\cal N} \int^1_0 \frac{dz}{2z} {\Big\langle {\langle \psi_0 |
{\cal H}^{\rm cmp}_{i;{\cal N}}+ {\cal H}^{\rm stu}_{i;{\cal N}}
| \psi_0 \rangle} \Big\rangle}_K[z\OV]
\cr
&=&
\int^1_0 \frac{dz}{2z} {\Bigl(
G:\Xi[z\OV;G]:G + G:\Upsilon[z\OV;G]:G
\Bigr)}
~~~ ~~~ 
\cr
&\equiv&
\Phi^{\rm cmp}[\OV] + \Phi^{\rm red}[\OV].
\label{irrX3.1}
\end{eqnarray}
The propagator is determined by Eqs. (\ref{irr03}) and (\ref{aux05}) with the
exact self-energy $\Sigma \equiv (\Xi + \Upsilon):G$. The kernels $\Xi$
and $\Upsilon$ are coupled by their joint renormalization of $G$.

Being exact, $\Gamma$ is complete to all orders in $V$. It supports every
permissible skeleton diagram, irreducible or not. Unlike a $\Phi$-derivable
approximation, no new two-body vertex can be generated that does not
already appear in
\begin{eqnarray}
\Gamma = \frac{\delta^2 \Phi}{\delta G \delta G}
&\equiv&
\Upsilon'[G] + \Xi'[G]
\cr
&=&
\Upsilon[G] + \Xi[G] + \Upsilon''[G] + \Xi''[G].
~~~ ~~~ 
\label{lam02}
\end{eqnarray}
The sum $\Upsilon'[G] + \Xi'[G]$ is naturally conserving while $\Xi'$ would
only be so autonomously with its own propagator fixed from
Eq. (\ref{irrX1.0}). However, $\Upsilon'$ is not independent of $\Xi$ and
so cannot stand as an autonomous conserving kernel although
$\Upsilon'[G] = \delta^2\Phi^{\rm red}/\delta G \delta G$ is well defined;
but now the exact nature of $\Gamma$ in Eq. (\ref{lam02}) implies
\begin{eqnarray*}
\Upsilon''[G] + \Xi''[G] = 0
\label{lam02.1}
\end{eqnarray*}
meaning, with $G$ exact, that mutual coupling of the two species of vertex
terms contrives to restructure the residuals such that what were previously
distinct noncrossing symmetric components are now absorbed into $\Gamma$ in
crossing symmetric partnership with their counterparts.

\subsection{Kraichnan parquet for the exact ground state}

Having recast the exact Hamiltonian in the Kraichnan embedding as a strictly
pair-linked plus a complementary part, we have split the LW functional into
the $stu$ irreducible $\Phi^{\rm cmp}$ so $\Phi^{\rm red}$ must then exhibit,
on an equal footing, all three possible modes for pair-reducible composite
excitations.
One can account explicitly for all permissible topologies, in all combinations
occurring in the exact kernel $\Gamma$,
by returning to Eqs. (\ref{kII15.5}) and (\ref{kII15.6}). Enlarge
Eq. (\ref{kII15.4}) to
\begin{eqnarray}
{\cal V}[\ophi] \equiv \OV  + \Upsilon'' + \Xi + \Xi''
= \OV  + \Xi[\OV \ophi];
\label{kII15.4b}
\end{eqnarray}
%
there are now no residuals. As with $\Lambda[\OV\varphi]$ earlier, we retain
the (otherwise redundant) anticoupling in $\Xi[\OV\ophi]$ to provide an audit
trail for the action of both $\ophi$ and $\varphi$ in the following.

Equation (\ref{kII15.4b}) includes all terms in the complete kernel that are
not $stu$ pair reducible overall. Nevertheless, examination of
Figs. \ref{F8}(b), (c) and (e) tells us that some diagrams in the
$stu$-irreducible expansion of $\Phi^{\rm cmp}$ are similar to the purely
$stu$ series of Fig. \ref{F8}(a) in having components separable by cutting
two pairs of propagators. Thus the expansion of $\Xi$, while excluding
$stu$ structures in the global sense, will still include convolutions
that are pair reducible. One could reverse engineer the structure of the
Kraichnan $\Xi[\OV\ophi]$ to identify its primitively irreducible terms,
such as in Figs. \ref{F8}(d) and (f). While not needed here,
the procedure to do so is in Appendix C.

The strategy for obtaining the complete $\Gamma$ now follows the same
methodology
\cite{pqt3}
of sorting out the individual $stu$ pairing channels as in
Eqs. (\ref{kII15.5}) and (\ref{kII15.6}). The extended Kraichnan parquet
equations for the exact ground state are
\begin{widetext}
\begin{eqnarray}
\Gamma_s
&=&
\OV + {\cal V}[\ophi] +{\phi}^{-1} (
  \Gamma\varphi :GG: t\Gamma_t
- \Gamma\varphi :GG: u\Gamma_u
);
\cr
\Gamma_t
&=&
{\cal V}[\ophi] + {\phi}^{-1} (
- \Gamma\varphi :GG: u\Gamma_u
+ \Gamma\varphi :GG: s\Gamma_s
);
\cr
\Gamma_u
&=&
{\cal V}[\ophi] + {\phi}^{-1} (
  \Gamma\varphi :GG: s\Gamma_s
+ \Gamma\varphi :GG: t\Gamma_t
)
\cr
\cr
{\rm with}~~
\Gamma
&\equiv&
\OV + {\cal V}[\ophi] + {\phi}^{-1} (
  \Gamma\varphi :GG: s\Gamma_s
+ \Gamma\varphi :GG: t\Gamma_t
- \Gamma\varphi :GG: u\Gamma_u
)
\label{kII15.A}
\end{eqnarray}
\begin{eqnarray}
{\rm so}~~
\Upsilon
&=&
2\OV + {\phi}^{-1} (
  \Gamma\varphi :GG: s\Gamma_s
+ \Gamma\varphi :GG: t\Gamma_t
- \Gamma\varphi :GG: u\Gamma_u
).
\label{kII15.B}
\end{eqnarray}
\end{widetext}
This four-point representation is not for
any two-body collision process whatsoever, but is tied to the particle-hole
dynamics dictating the system response to an external, number-preserving
probe. The overall outer coupling ${\phi}$ still has to be carried on the
right-hand side of Eq. (\ref{kII15.A}) as the essential bookkeeping device
to that end. By the principle of Fig. \ref{F7}, the autonomous scattering
elements linked by $GG$ pairs in the equations above have cycles of
K couplings or anticouplings that factor out to unity independently.

\subsection{Standard parquet and the exact ground state}

We now address the formal distinction between (a), the rather different
specification of parquet via the exact Eqns. (\ref{kII15.A}) and
(\ref{kII15.B}), and (b) the standard parquet version in which K couplings
and anticouplings do not exist and the distinction among channels is
made purely through the particular momentum-flow combinations of the three
possible $GG$ pairings in $stu$ 
\cite{k2};
no further knowledge is adduced here to establish which irreducibles are
represented in $\Xi$. Unlike in the embedded-Hamiltonian approach, there is
no recipe to identify explicitly the content of the (conserving) $stu$
subseries and, importantly, that of its complement.

For case (a) we recall how the (relative) causal ordering of the
propagators is chosen automatically. Equation (\ref{irr01}) for the
Hamiltonian fixes {\em ab initio}, through the definition of $\varphi$ in
Eq. (\ref{irr02}), how the creation-annihilation operators are to couple.
For (b) this is done by imposing the three possible
dynamical flows on the form of the parquet equations. In Kraichnan
the parquet structure emerges more naturally.

The exact Hamiltonian and $\Phi$ are independent of the K couplings.
There are two obvious choices for $\varphi$.
\begin{itemize}
\item
Choosing $s = t = u = 1$ means $\ophi = 0$. Then
${\cal H}^{\rm cmp}$ and $\Xi$ vanish so
$\Upsilon = \Gamma$. Since access to the innermost structure of $\Gamma$ is
unavailable, Eqs. (\ref{kII15.A}) and (\ref{kII15.B}), though exact, hide
the essential contributions of the irreducible versus the strictly pairwise
correlations. No insight is gained.
\item
The choice $s = t = u = 0$ forces $\ophi=1$ and now ${\cal H}^{\rm stu}$ and
$\Upsilon$ are zero in Eq. (\ref{irrX3.1}). Then $2\OV+\Xi$ becomes $\Gamma$,
carrying everything for the exact problem while $\Upsilon$ in
Eq. (\ref{kII15.B}) goes to $\Upsilon-2\OV$ which vanishes at order
$\varphi$. Once more there is no gain.
\end{itemize}
It would be inconsistent to set $\ophi=\varphi=1$
in Eq. (\ref{kII15.A}) ignoring the anticorrelation in $\Xi[\OV(1-\varphi)]$
actuated through the Kraichnan phase average. The invariance of the exact
Hamiltonian, Eq. (\ref{irrX1.1}), would be violated. Choosing $\ophi$ and
$\varphi$ as unity in Eq. (\ref{irrX1.1}) simply doubles the interaction
strength, which is not equivalent to putting $\ophi=\varphi=1$ in
Eqs. (\ref{kII15.A}) and (\ref{kII15.B}) since the interaction $\OV$ there
no longer matches $2\OV$ in the rescaled Hamiltonian.

As already pointed out, if an exact expansion of $\Gamma$ were to dispense
with a classification such as Kraichnan coupling, an extra intervention would
be needed to avoid redundant graphs in the progression from $\Gamma$ up to
$\Phi$, Moreover, without a way to discriminate systematically between
pairing and irreducible sectors, the irreducibles in $\Gamma$ must be picked
out basically by inspection.

\subsection{Rationale for the Kraichnan approach}

In the exact case the question is: doesn't the Kraichnan construction merely
replicate, by rather more convoluted reasoning, what standard parquet
already conveys? The answer is no, and it comes in two parts. The first
concerns the different way in which the standard treatment couples the
pairing channels in its analog to $\Upsilon[\OV]$, Eq. (\ref{kII15.B}).
The second is the fate of the unavoidable truncations of the exact expansion
within the respective parquet accounts.

\begin{figure}
\centerline{
 \includegraphics[height=6.5truecm]{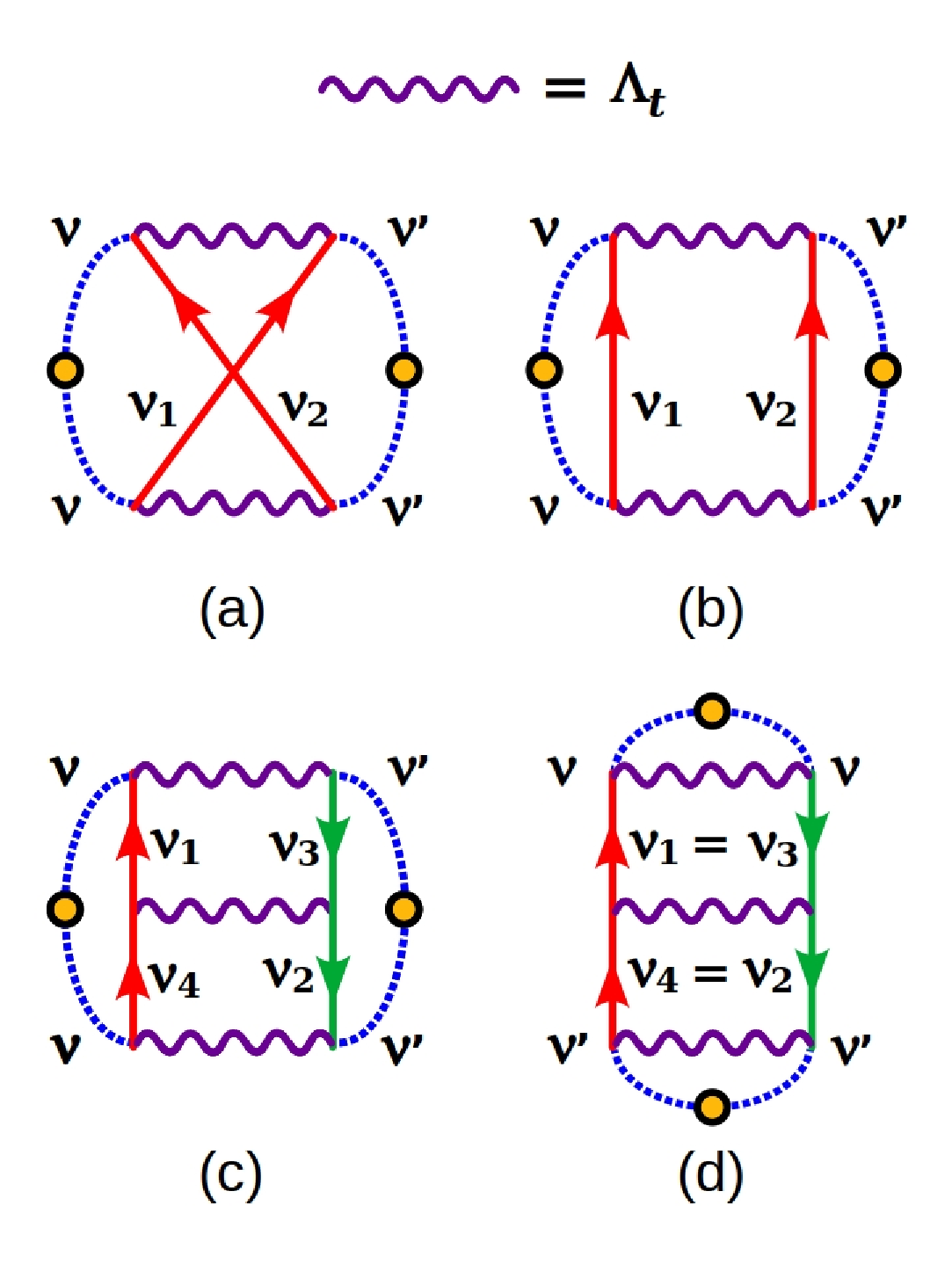}}
\caption{
Composite scattering terms presented within their response contributions.
(a) $s$-channel process, mediated by the screened interaction $\Lambda_t$
from Fig. \ref{F3}(a) and inadmissible as a $stu$ term, unlike its direct
counterpart (b), corresponding to $\Lambda_{t;s}$ in Fig. \ref{F4}(b).
(c) $u$-channel ladder again mediated by $\Lambda_t$, also $stu$
inadmissible, unlike (d). Figures (a) and (c) are admissible when $\OV$
replaces $\Lambda_t$, as the first survives in pure $s$-channel mode
and the second in pure $u$-channel mode. At higher orders in $\OV$, the
$t$-coupling phases associated with $\Lambda_t$ cancel only in negligible
ranges of their indices in the limit of infinite system copies. When
mediated by their anticouplings, however, such non-$stu$ diagrams survive
and appear in the complementary series generated by $\Phi^{\rm cmp}$.
}
\label{F10}
\end{figure}

Figure \ref{F10} illustrates the issue for pairing-channel diagrams typically
met in conventional parquet, alongside their exchanges which are common
to it and $stu$. Although legitimate Feynman terms
\cite{pqt2},
Figs. \ref{F10}(a) and (c) are excluded from the $\Phi$-derivable $stu$
model. For Fig. \ref{F10}(a) the total K coupling phase over the pair
of $t$-channel interactions is subject to $s$-like exchange indexing.
Following Eq. (\ref{irr02}) it becomes
\[
\tau_{\nu\nu_2} + \tau_{\nu'\nu_1} + \tau_{\nu_2\nu'} + \tau_{\nu_1\nu}
=
\tau_{\nu\nu_2} - \tau_{\nu\nu_1} + \tau_{\nu'\nu_1} - \tau_{\nu'\nu_2}  
\]
and vanishes only for the asymptotically negligible ranges $\nu=\nu'$ or
$\nu_1=\nu_2$. At the same time the phases for Fig. \ref{F10}(b),
replicating $\Lambda_{t;s}$ from Fig. \ref{F4}(b), cancel identically so
$\Lambda_{t;s}$ is $stu$ admissible. A similar situation holds for
Fig. \ref{F10}(c) vis \`a vis the $stu$ term \ref{F10}(d): the latter is
admissible, its exchange is not.

\vskip 0.15cm

In contrast with the above, the standard parquet summation convention for
pairing channels in the kernel $\Gamma$ takes the contents of
Eq. (\ref{kII15.A}) with no coupling factors. Consequently it is blind to
the source and destination of the $GG$ pairs on either side of every
interaction $V$.

Standard parquet assigns diagrams to the pairing channels of its $\Upsilon$
analog that cannot appear in the corresponding Kraichnan object. The exact
expansion executed in standard parquet analysis requires a version of the
irreducible $\Xi$ that is a subset of the skeletons incorporated in the
Kraichnan $\Xi[\OV\ophi]$ associated with Eq. (\ref {irrX1.0}), because, as
with the examples in Fig. \ref{F10}, structures that count as pair reducible
in the standard formulation are excluded in the $stu$ formulation. The
irreducible seed kernel required by parquet analysis corresponds to the one
identified in Appendix C.

As long as one considers the exact ground state, there seems little to choose
between the two versions of parquet, but that is not the case. In the version
proposed here, the equations emerge from a Hamiltonian, informed by the
constraints of unitarity as well as from a more natural partitioning of
correlations into $stu$ and complementary sectors. In their conventional form
the parquet equations are inferred bottom-up; for the exact case their
structure presumes the existence of a well defined set of irreducible kernel
components. Nevertheless it lacks a systematic way to construct the full set.
Passage to Kraichnan anticouplings provides an explicit selection algorithm,
at least in principle.

The conceptual difference in codifying correlations has practical
consequences. Since it is necessary to truncate the exact expansion for a
viable calculation, in Kraichnan's version of $\Phi$ derivability one can do
so by selecting a physically motivated subset of the irreducible complement
$\Phi^{\rm cmp}$. While such an essentially intuitive choice means loss of
contact with a formal Hamiltonian and with crossing symmetry, the generic
Luttinger-Ward structure of Eq. (\ref{irrX3.1}) persists to sustain
selfconsistency and conservation; the Kraichnan analysis falls back to
Baym and Kadanoff
\cite{kb1,kb2}.
The price of retaining crossing symmetry in the standard parquet equations,
on the other hand, is loss of contact with unitarity
\cite{roger}.

It is worth revisiting a related aspect with regard to response. There,
the tight coordination between self-energy corrections carried in $G$, or
``out-of-the-beam'' scattering, and two-body scattering vertices, or
``into-the-beam'' scattering, is the crucial element in conservation
when multi-pair excitations are involved
\cite{qsm}.
The functional definition $\Lambda' = \delta \Sigma/\delta G$ underpins
the approximate but conserving Bethe-Salpeter equation. It is then essential
for the extra noncrossing symmetric vertex terms to be kept in $\Lambda'$.
If the parquet kernel $\Gamma$ is to remain crossing symmetric by discarding
the incompatible terms that would come from variation of its own associated
self-energy $\Gamma:G$, the conserving nexus between $\Sigma$ and $\Gamma$
breaks down, compromising the response properties.

The inference from this Section is that while the diagrammatic structure of
the ground state uniquely conditions the exact conserving parquet equations,
the traditional parquet equations, set out heuristically, cannot by themselves
automatically reconstitute the ground state, nor the excitations defining the
physics of response. Additional correlational knowledge is needed.

\section{Completeness, Uniqueness, Consistency}

\subsection{Completeness and crossing symmetry}

We are ready to advance a Kraichnan-based clarification of the results of
Refs.
\cite{becker,js,roger}
foreclosing all chance of incorporating both crossing symmetry and
conservation in any truncated description of an interacting system.
For this we return to the basic definition that embeds the physical
interaction Hamiltonian in the Kraichnan collective form renormalized by
its K couplings:
\begin{eqnarray}
{\cal H}_{i;{\cal N}}[V \varphi]
&=&
\frac{1}{2{\cal N}} 
{\sum_{k_1 k_2 k_3 k_4}}\!\!\!\!' ~~~{\sum_{\nu_1 \nu_2 \nu_3 \nu_4}}\!\!\!\!'
{\langle k_1 k_2 | V  | k_3 k_4 \rangle}~
\cr
&&
~~~ ~~~ 
\times
\varphi_{\nu_1\nu_2|\nu_3\nu_4}~
a^{\dagger}_{\ell_1} a^{\dagger}_{\ell_2} a_{\ell_3} a_{\ell_4}
\label{iv01}
\end{eqnarray}
in which, for clarity, we resolve the index $l = (k,\nu)$ into its physical
and collective components and we leave unsymmetrized the elementary
interaction.

Recall that the Luttinger-Ward functional from Eq. (\ref{iv01}) is exact for
this specific Hamiltonian. The corresponding kernel $\Gamma[V\varphi]$ is
unique, conserving and crossing symmetric. What happens, then, in the average
over the stochastically defined K couplings? Consider the collective Fock
space of a typical member of the ${\cal M}$-sized superensemble, with
distribution $\{\varphi\}$ and Hamiltonian ${\cal H}_{i;{\cal N}}[V\varphi]$.

When $\varphi \equiv 1$ we have a direct sum of physically identical but
distinguishable Fock spaces arranged, as it were, as ${\cal N}$ block
diagonals for which the collective description indexed by $\nu$ has no
operative role, and the expectation for $\Phi[V]$ over the discrete blocks
is exact. As soon as the K coupling function takes a nontrivial form, there
is cross-linking among the Fock-space copies making up the collective set;
correlations are induced across copies, off the block-diagonal and mediated
by the indexed $\varphi$.

Any closed cross-copy interaction graph has a topology identifiable within
the exact physical expansion, with the addition of the spin-like indices
$\nu$. It must have an exchange counterpart also diagrammatically identical
to the physical exchange. Thus crossing symmetry applies to the pair. The
presence of $\varphi_{\nu_1\nu_2|\nu_3\nu_4}$ does not in itself destroy
formal crossing symmetry in $\Gamma[V\varphi]$, but the exchange of indices
scrambles the overall K coupling phase in a very different fashion from how
physical exchange acts on the momentum transfer for $V$.

In $stu$, composite correlations in the form of Fig. \ref{F6} or \ref{F7}
numerically survive the stochastic phase average but are no longer guaranteed
to have crossing symmetric exchange partners that also survive. Their
exchanges, as with Figs. \ref{F10}(a) and (c), will have zero Kraichnan
phase only in an asymptotically negligible range over the index
representation. Meanwhile the total Kraichnan phase for the allowed terms,
as with Figs. \ref{F10}(b) and (d), will be identically zero over the entire
range of collective indices. The forward-scattering sum rule
\cite{tom,vll},
that is ${\langle kk|\Lambda|kk \rangle} = 0$, is violated for the $stu$
residuals $\Lambda''$ even if not for $\Lambda$. This is one instance of
the loss of the completeness (coherence) of the extended Fock space; another
is discussed below for the structure factors.

Stochastic averaging kills the off-diagonal exchange correlations in the
collective index description of the cross-coupled Fock-space blocks,
thus also the (still formally present) crossing symmetry. However,
conservation, including for the $\Phi$-derivable dynamical response
structure, depends only on the properties of the surviving index-diagonal
correlations. They are fixed by the recursive topology of the
originating unitary Luttinger-Ward functional, invariant under perturbations.

Prior antisymmetrization of $V$ at the Hamiltonian level ameliorates the
loss of exchange symmetry by making it explicit at the individual
interaction level. This is not sufficient to secure global crossing
symmetry. In general it is not local topology but that of entire sets
of diagrams in superposition, that determines crossing symmetry for the
complete assembly; a principle that also applies to conservation within its
sector.

In terms of the extended Fock-space scenario, one might think of parquet
theory as the partial importation of off-diagonal correlation structures in
index space, namely the exchange complements, to act side by side with
diagonal conserving ones. When done in a way that is uncontrolled
{\em from the Kraichnan perspective} at any rate, it is not surprising that
conservation is compromised.

\subsection{Uniqueness and structural ambiguity}

There remains the inevitable ambiguity in defining the effective kernel
$\Lambda$ for a Kraichnan model LW functional. Antisymmetrization of the
elementary potential makes it easier to construct a physically reasonable
crossing symmetric seed kernel, but one could more crudely restructure the
closed graphs of $G:\Lambda:G$ by directly antisymmetrizing $\Lambda$.
We have seen that the Kraichnan average must decohere crossing symmetric
superpositions while maintaining conservation. The variation
$\Lambda'=\delta^2\Phi/\delta G \delta G$ is then bound to produce
additional terms $\Lambda''$ distinct from $\Lambda$ and that cannot be
crossing symmetric, else $\Lambda'$ would be crossing symmetric and
conserving, so $\Lambda$ itself would be unique, being exact.

As shown in Appendix A, residual terms are able in their own right to
reconstitute $\Phi$ by reclosing with two propagators and integrating the
Hellmann-Feynman identity in Eq. (\ref{lw47.1}).
This in turn means that $\Lambda$ need not be unique as a generator
for the LW functional.
A concrete example for the $stu$ model would be to sum appropriate terms in
Fig. \ref{F5}, for instance:
\begin{itemize}
\item[]
$\Lambda_{s;t}$ from the $s$-channel, Fig. \ref{F5}(a),
\item[]
$\Lambda_{t;u}$ from the $t$-channel, Fig. \ref{F5}(b), and
\item[]
$\Lambda_{u;s}$ from the $u$-channel, Fig. \ref{F5}(c).
\end{itemize}
In passing we note that there are identities between the extra kernel terms
in Fig. \ref{F5}, not explored here, that render the outcome of bosonic
Kraichnan different from the present fermionic one.

\subsection{Two-body Consistency}

\begin{figure}
\centerline{
 \includegraphics[height=5truecm]{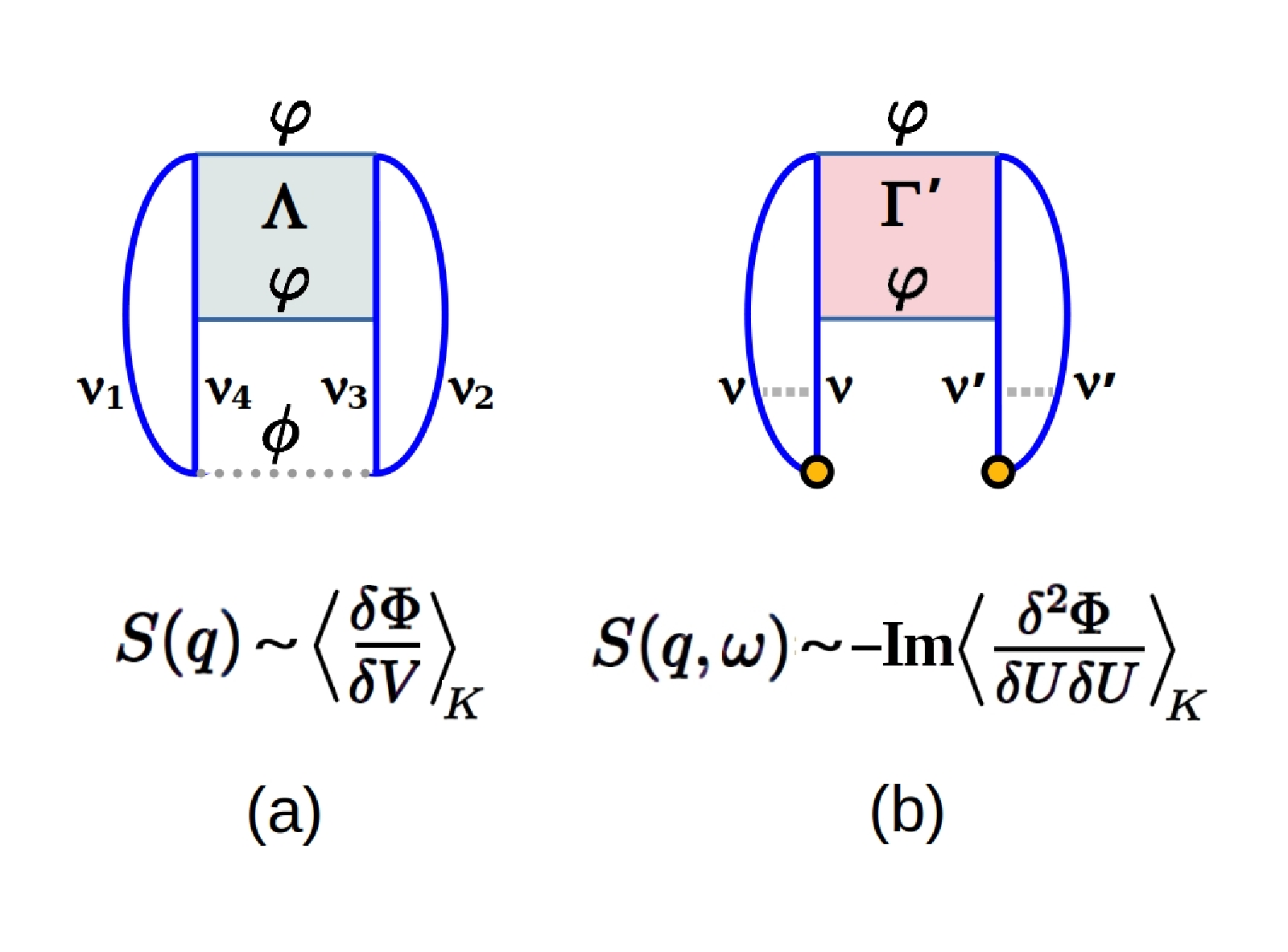}}
\caption{
Two aspects of two-body diagrammatics, contrasted.
(a) Generation of the static structure factor $S(q)$ through removal of one
interaction line from $\Phi$. In the $stu$ model the original K coupling
$\phi$, concomitant with $V$, is unaffected. It stays convolved with the
internal couplings $\varphi$ through the collective indices. A Kraichnan
averaging yields the primary $stu$ kernel $\Lambda$ of Fig. \ref{F3}(a), defining
$S(q)$ from Eqs. (\ref{aux02}) and (\ref{irr03x}).
(b) Generation of the dynamic structure factor $S(q,\omega)$ under an
external perturbation. The collective-index
rules allow for a larger set of terms in the response of the total kernel
$\Gamma'$, Eq, (\ref{kII18.4}). In the exact expansion
$\Gamma' = \Gamma = \Lambda$ and $S(q)$ is identical to the frequency
integral of $S(q,\omega)$ in Eq. (\ref{aux04}). In a $\Phi$-derivable
approximation (b) does not replicate (a). In parquet theory, $\Gamma$ is
iterated heuristically to redefine $\Lambda$
\cite{pqt3},
bootstrapping the self-energy $\Gamma:G$ until convergence. This still
cannot result in a conserving $S(q,\omega)$ even if its (a) and (b) analogs
were compatible.
}
\label{F11}
\end{figure}

We have explored the $stu$ model in its Kraichnan setting through
the single-particle formalism of Kadanoff and Baym
\cite{kb1,kb2}
in the language of Luttinger and Ward
\cite{lw}.
To probe $\Phi$ derivability from a perspective closer to the object of
parquet theory, we switch the topic to two-particle processes as they fix
the structure factors and the status of Eq. (\ref{aux04}) in Sec. II.
There we reviewed the role of the two-body Green function in
structuring the exact Luttinger-Ward functional; here we take it up
in the setting of a conserving approximation. This leads to reevaluation
of the relation between static (more properly, equal-times) and dynamic
response functions
\cite{dp}.

The primary definition of $S(q)$ is given by Eq. (\ref{aux02}), not
Eq. (\ref{aux04}) which is a consequence of Fock-space completeness.
Here we recapitulate its perturbative relation to the exact kernel
$\Gamma$. From Eq. (\ref{lw47.1}) in Appendix A we obtain the
variational derivative, with the full physical propagator $G[V]$
in contrast with Eq. (\ref{aux05.1}):
\begin{eqnarray*}
\frac{\delta \Phi}{\delta V}
&=&
(I - G^{(0)}\!\cdot\!\Sigma~\cdot)^{-1}
\cdot G^{(0)}\cdot \frac{\delta \Sigma}{\delta V}
\cr
&&
-~ G:\!\frac{\delta \Sigma}{\delta V}
- \frac{\delta G}{\delta V}:\Sigma
+ \frac{\delta \Psi}{\delta V}
\cr
\cr
{\rm where}~~
\frac{\delta \Psi}{\delta V}
&=&
\frac{\delta G}{\delta V}:(\Gamma:G)
+ 
G:\frac{\delta \Gamma}{\delta V}:G.
\end{eqnarray*}
Cancellation of terms results in
\begin{eqnarray*}
\frac{\delta \Phi}{\delta V}
&=&
G:\frac{\delta \Gamma}{\delta V}:G
\end{eqnarray*}
and the right-hand expression can be inserted into Eq. (\ref{aux02})
to obtain $S(q)$ as schematized in Fig. \ref{F11}(a).
For a $\Phi$-derivable model
with an approximate $\Lambda$ replacing $\Gamma$, the proof
follows through. Moreover, the static structure factor is real-valued
\cite{dp,mahan2}
since analyticity (the closed diagrams' causal structure) is unaffected.

At least in principle, constructive parquet theory builds up its own
irreducible $\Xi$ by successively inserting an additional link $V:GG$ within
the self-energy $(\Gamma\!-\!V):G$ to replace it with $(V:GG:\Gamma):G$
avoiding overcounting of terms already generated via the pairwise parquet
equations; this generates new primitive irreducibles alongside the preceding
input expression. Then $\delta \Sigma/\delta G$ is recalculated. Old and new
crossing symmetric terms are retained to form a new $\Xi$. Noncrossing
symmetric terms, as unavoidable here as in $\Lambda'$ for $\Phi$
derivability, are discarded. The parquet equations are solved once again
for a new $\Gamma$. Details can be found in Bickers
\cite{pqt3}.

In parquet, by the logic above, iteration leads to $\Lambda$ and $\Gamma'$,
or more properly $\Gamma$, converging to the same crossing symmetric object
in both Fig. \ref{F11}(a) and \ref{F11}(b). Without connection to a
variationally consistent, conserving two-body Green function, it is unclear
whether its $S(q,\omega)$ would accord with $S(q)$ in Eq. (\ref{aux04}).

From an entirely different cause, violation of Eq. (\ref{aux04}) is a known
characteristic of $\Phi$ derivability. In this case, the reason is the
loss of correlated-state coherence in Kraichnan averaging
\cite{KII}.
A significant instance of this violation is the random-phase approximation,
for which $S(q)$ is always (trivially) positive while its real-space static
pair correlation function, obtained from integration of $S(q,\omega)$,
becomes negative in the short-range limit
\cite{mahan2}.

Time translation invariance of the $\Phi$-derivable ground-state description
implies that if $\Lambda$ were to replace $\Gamma'$ in Fig. \ref{F11}(b)
to define an object that we may call $S_{\Lambda}(q,\omega)$, then trivially
its inverse Fourier at equal times is
\begin{eqnarray*}
S(q)
&=&
\frac{1}{N} \int^{\infty}_0 d\omega S_{\Lambda}(q,\omega).
\end{eqnarray*}
Similarly, if we replace $\Lambda$ in Fig. \ref{F11}(a) with $\Gamma'$
and call this object $S_{\Gamma'}(q)$, then
\begin{eqnarray*}
S_{\Gamma'}(q)
&=&
\frac{1}{N} \int^{\infty}_0 d\omega S(q,\omega).
\end{eqnarray*}
Consistency in a $\Phi$-derivable calculation precludes confusing
$S_{\Gamma'}(q)$ with the proper $S(q)$ defined as indicated in
Fig. \ref{F11}(a), and confusing $S_{\Lambda}(q,\omega)$ with the proper
$S(q,\omega)$ defined in accordance with Fig. \ref{F11}(b).

\section{Summary}

In this work we have returned to the basic makeup of diagrammatic
expansions for the strongly interacting ground state, offering a different
understanding of the puzzling incompatibility between two classic many-body
methods: parquet and $\Phi$-derivable approximations. The first cannot
sustain conservation for its response functions and the second cannot
sustain crossing symmetry for its complete two-body scattering kernel.
To cast another light on the established proofs of this incompatibility
\cite{becker,js,roger}
we have advanced an interpretation via Kraichnan's stochastic Hamiltonian
embedding.

Resolution of the exact Hamiltonian with this machinery lets one draw the
following distinction between the approaches in question. On the one hand the
conventional analysis of the parquet equations accommodates, in a more
intuitive way, a range of inter-channel pair correlations beyond those from
the Hamiltonian-based $stu$ formalism we have described. On the other hand,
from the Kraichnan derivation of exact parquet, there emerges a systematic
algorithm to isolate, in theory, all the pair-irreducible terms in a rational
way faithful to the exact Hamiltonian template.

Tracking the functional interaction between pair and irreducible processes
in fine detail seems beyond the current scope of either analysis. It is hard
to avoid the suspicion that crossing symmetry in the exact Luttinger-Ward
functional is underwritten by just such interplay. Separating out pair
processes, to study them in minimal configurations, has long accounted for
much crucial physics
\cite{a+m},
with great success. Beyond this remarkable record, clarifying further effects
is likely to bring in more than two-body dynamics. While these remain to be
explored beyond more refined mean-field pictures
\cite{tom,dmft},
theoretical investigations of three-body parquet already exist
\cite{lande}.

The difference between parquet and Kraichnan philosophies can be viewed in
analogy with two railroad switchyards for which, in the former case,
decisions as to which units couple where are primarily made locally at each
junction while, in the latter, they are primarily made globally and
autonomously by a central algorithm. A microscopic model ideally respects
the global topology that guarantees unitarity, or conservation, for the
associated response behavior
\cite{kb1,kb2,qsm}.
$\Phi$ derivability ensures this, but only at the price of losing the
crossing symmetry characteristic of fermionic systems.

Developments based on Kraichnan's approach might go in several directions.
In the first place, a broader study of the range of dynamical sum rules for
idealized models, such as Hubbard, would put practical numbers on rates and
kinds of violation by parquet and $stu$-FLEX treatments of the same example.

Kraichnan's original papers
\cite{k1}
and especially
\cite{k2}
provide straightforward stability proofs for the bound states of the ladder
($s$ channel only) and ring ($t$ channel only) models. While he foreshadows
multichannel extensions such as the present $stu$ picture, a corresponding
stability analysis for $stu$ is not at hand. The fact that it involves a
nontrivial superposition of the elementary channels complicates matters,
offering a natural topic for further work.

An issue of relevance is the examination of the boundary conditions
assumed in parquet over against $\Phi$-derivable theories which, at face
value, are very different. Parquet, conceived for strong scattering in open
systems, relies on exact particle-antiparticle correspondence. In adapting
parquet to condensed matter, one has then to consider: (a) that
(quasi)particle antisymmetry holds strictly only at the Fermi surface;
(b) that the system is closed to particle entry and escape; thus (c) that
"incoming/outgoing" particles in the condensed state are not asymptotically
free of the collective background. Conversely, conserving models address the
ground state and relatively low-lying excitations (more generally the free
energy). It does not then follow that they are appropriate to open-system
scattering from/to the unbound vacuum.

Reexamination and development of Kraichnan's technique could motivate
exploring other contexts in which the approach may be relevant, namely
those reliant on a Hamiltonian or that can be referred back to one. In
particular, for bosonic systems the additional kernel terms represented
by $\Lambda''$ no longer clash with particle symmetry. Then the distinction
between $\Phi$ derivability and parquet really comes down to each one's
relationship to conservation.

\section{Acknowledgments}

I thank T. L. Ainsworth, M. P. Das and R. A. Smith for many enlightening
discussions, over a long period, on the nature of many-body diagrammatic
expansions. Their reflections have stimulated my revisiting the topics
discussed in this and earlier papers.

\appendix

\section{Self-energies with less than full symmetry}

In this Appendix we recall general criteria for the status of closed diagrams
contributing to the Luttinger-Ward functional at finite order in the
interaction. The discussion is simplified by applying the equivalent
expression for the LW correlation energy functional, due to
Luttinger and Ward
\cite{lw}:

\begin{eqnarray}
\Phi[\OV;G]
&=&
- {\rm Tr}\{ \ln(I - G^{(0)}\!\cdot\!\Sigma~\cdot) \}
- G[\OV]\!:\!\Sigma + \Psi[\OV;G];
\cr
\Psi[\OV;G]
&\equiv&
\! \int^1_0 \frac{dz}{2z} G[\OV]\!:\!\Gamma[z\OV;G[\OV]]\!:\!G[\OV].
\label{lw47.1}
\end{eqnarray}
Here the one-body propagators $G$ are everywhere renormalized with their
internal interaction at full strength. Unlike the coupling-constant integral
in Eq. (\ref{irr03}), in the above $\Psi[\OV;G]$ covers only the multiplicity
of the interactions within the skeleton for $\Gamma$.

On the basis of the self-energy's being given uniquely by $\Sigma=\Gamma:G$,
variation with $G$ of $\Psi$ in Eq. (\ref{lw47.1}) again returns the
result of Eq. (\ref{aux05.1}), but for tracing combinatorics within
the LW skeletons, the diagrammatic expression in the form of $\Psi$
is more straightforward than its correlation energy expression $\Phi$.
One need only consider each graphical contribution to $\Psi$ avoiding
the implicit $\OV$ dependence of $G$.

For a given contribution to $\Psi$ at order $n$ in $\OV$ it is clear that,
if it belongs to $\Phi^{\rm stu}$ as in Figs. \ref{F3}(b) and \ref{F8}(a), or
else if absolutely irreducible as, for example, in Figs. \ref{F8}(d) and (f),
every propagator in such a diagram is strictly equivalent to every other,
yielding $2n$-fold symmetry. The combinatorial weighting $(2n)^{-1}$, carried
in the evaluation of the term, is duly undone in the variation
$\delta \Psi/\delta G$ which is the self-energy.

\vskip 0.25cm
\begin{figure}
\centerline{
 \includegraphics[height=6truecm]{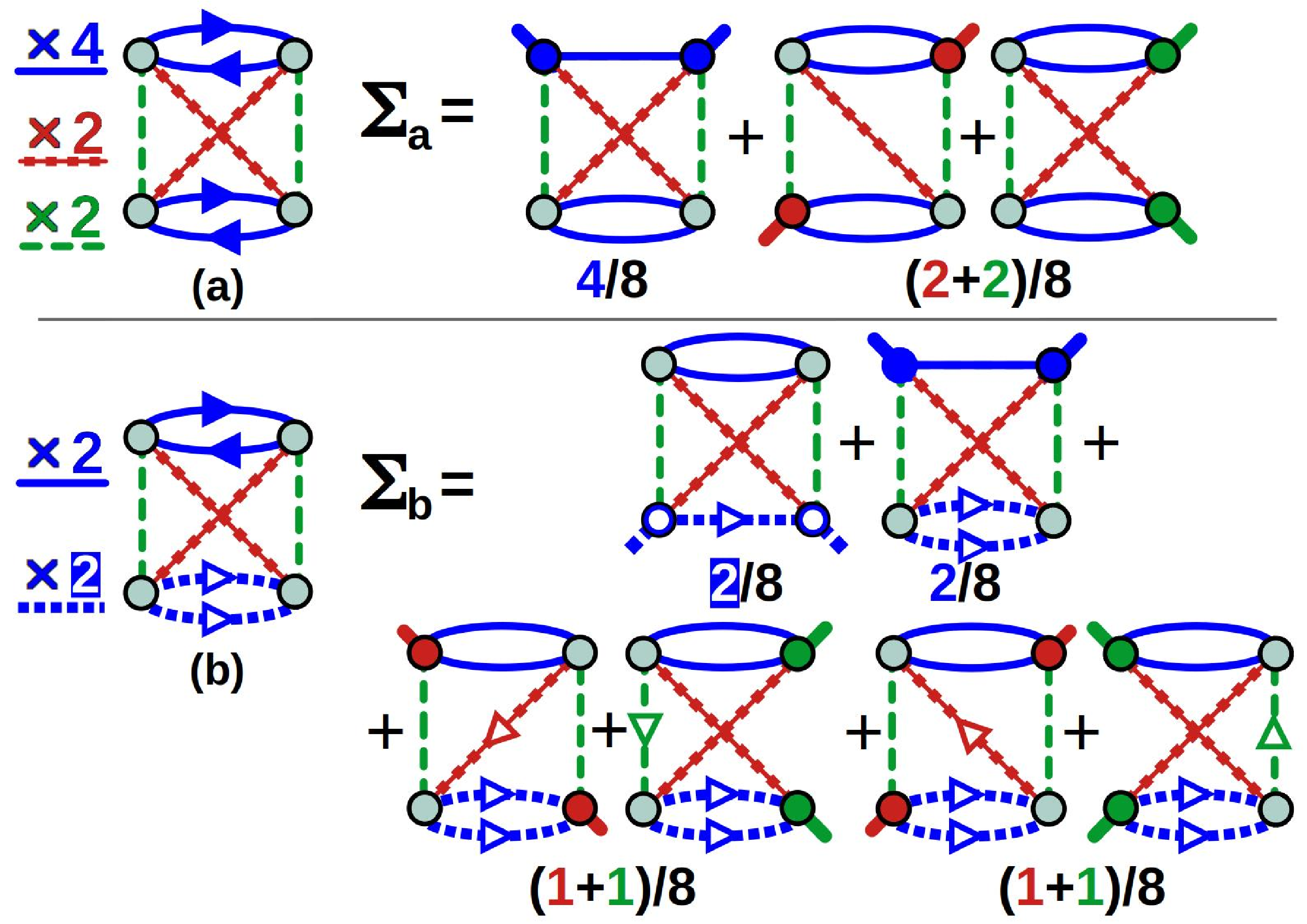}}
\caption{
Leftmost graphs: two variants of the fourth-order
non-$stu$ contribution to the LW functional, after Fig. \ref{F8}(b), whose
structural symmetry is lower than the eightfold maximum.
(a) Right-hand diagrams: variation for the self-energy
$\Sigma^{[4]} = \delta \Psi^{[4]}/\delta G$ for the term with two
embedded polarization bubbles, yields one fourfold and two twofold
self-energy parts with appropriate weightings. (b) Right-hand diagrams:
$s$-type ladder replaces one $t$ bubble. Symmetry is lower than for (a) while
the total self-energy retains proper overall weighting. In each case the
replacement $\Gamma^{[4]}[z\OV;G]:G \to \Sigma^{[4]}[z\OV;G]$ in
Eq. (\ref{lw47.1}) exactly recovers $\Psi^{[4]}$; but so will any one of the
distinct self-energy components when closed up again by a $G$ line and again
introduced, with unit weight, as the coupling-constant integrand in
Eq. (\ref{lw47.1}). More widely, although the self-energy from any set of
closed skeleton diagrams is always derivable consistently and unambiguously,
a unique kernel $\delta \Sigma/\delta G$ may not be definable.
}
\label{F12}
\end{figure}

This is not so for composite irreducible diagrams such as those of
Figs. \ref{F8}(b), (c) and (e), which have less than complete $2n$-fold
symmetry. Since they carry inequivalent G lines, the way that a unique term
in $\Sigma$ is generated is less obvious. We consider the concrete example of
the fourth-order diagram Fig. \ref{F8}(b). Call its kernel
$\Gamma^{[4]}[\OV;G]$. After integrating out the coupling constant, its
contribution to $\Psi$ is
\begin{eqnarray*}
\Psi^{[4]} = \frac{1}{8} G:\Gamma^{[4]}[\OV;G]:G.
\label{lw47.2}
\end{eqnarray*}
In Fig. \ref{F12} we display the outcome of taking the variation with
respect to $G$ by opening up lines everywhere in $\Psi^{[4]}$. Given that its
diagrammatic symmetry is lower, variation with $G$ on each side generates
a set of dissimilar structures in the self-energy, but each comes weighted
by its symmetry factor, defining the total self-energy as their weighted sum.

The test for $\Phi$-derivable consistency is whether performing a
coupling-constant integral with the functional $\Sigma[z\OV,G]$ replacing
the form $G:\Gamma[z\OV;G]$ in the integrand on the right-hand side of
Eq. (\ref{lw47.1}) recovers the original fourth-order piece of the LW
functional. With the assigned symmetry weightings, it does. So
$\Psi^{[4]}$ and hence $\Phi^{[4]}$ are legitimate.

For a general closed graph of $\Phi$ of order $n$ in the interaction, the
procedure is straightforward. Let there be $k$ groups of topologically
interchangeable one-body propagators, each with $\gamma_i$-fold symmetry so
$\sum^k_{i=1} \gamma_i = 2n$. Then $\gamma_i$ should be even if every graph
is at least microscopically reversible. We can define a $\Psi^{[n]}$
accordingly as the corresponding component in Eq. (\ref{lw47.1}).
With $G_i$ representative of the $i$th group,
\begin{eqnarray}
\Sigma^{[n]}
&=&
\sum^k_{i=1} \frac{\gamma_i}{2n} \Sigma^{[n]}_i;~~  
\Sigma^{[n]}_i
\equiv \frac{\delta \Phi^{[n]}}{\delta G_i}.
\label{lw47.3}
\end{eqnarray}
Now any $\Sigma^{[n]}_i$ on the right-hand side of Eq. (\ref{lw47.3}),
given full unit weight, is enough on its own to reconstitute the full
structure when closed up again with a propagator so
$\Psi^{[n]} = (2n)^{-1}G:\Sigma^{[n]}_i[\OV;G]$. It follows that in
constructing a general component to approximate the LW functional, the
choice of an effective two-body scattering kernel $\Gamma$ may not be unique
(apart from first order, namely Hartree-Fock). This is clear for closed
graphs of less than maximal symmetry; but Fig. \ref{F2} for the second-order
self-energy and inspection of the diagrams making up $\Phi^{\rm stu}$
itself, Fig.
\ref{F8}(a),
show this to apply as well to $\Gamma$ for at
least a subclass of skeletons with maximal symmetry at all orders.
This is demonstrated in Fig. \ref{F5} for $stu$. 

\section{Kraichnan average of a skeleton graph}

Take any closed skeleton graph in the series for $\Phi[\OV\varphi]$
at order $n$ in $\OV$. Recalling that $\varphi = 1 - (1-s)(1-t)(1-u)$, the
Kraichnan average of the $n$ factors will be
\begin{eqnarray}
{\Big\langle \prod^n_{i=1} \varphi_i \Big\rangle}_K
&=&
1 - {\Bigl(1 - {\Big\langle \prod^n_{i=1} s_i \Big\rangle}_K \Bigr)}
\cr
&& \times
    {\Bigl(1 - {\Big\langle \prod^n_{i=1} t_i \Big\rangle}_K \Bigr)}
    {\Bigl(1 - {\Big\langle \prod^n_{i=1} u_i \Big\rangle}_K \Bigr)}.
~~~ 
\label{apx1}
\end{eqnarray}
To show this, consider a typical product in the expansion of the left-hand
side of Eq. (\ref{apx1}). It has the form
\begin{eqnarray}
{\Big\langle \prod^{[n_s]}_i s_i \Big\rangle}_K  
{\Big\langle \prod^{[n_t]}_j t_j \Big\rangle}_K  
{\Big\langle \prod^{[n_u]}_k u_k \Big\rangle}_K
\label{apx2}
\end{eqnarray}
where $[n_c]$ for each channel $c$ denotes that the product has $n_c$
factors with $0 \leq n_c \leq n$ and the stochastically uncorrelated
phases for each channel decouple in the overall average.

If an expectation for channel $c$ in the expression (\ref{apx2}) does not
vanish, it must be identically unity. Then its graph may be closed by
detaching it from the other channel products and thus represents a
legitimate, autonomous diagram of order $n_c$. However, if $n_c < n$ the
constraints on index sums means that one, and only one, pair $GG$ could have
connected the parts, similarly to Fig. \ref{F7}; connection by multipairs
could not force equality of the indices across every linking pair. Therefore
the subgraph must constitute a self-energy insertion and the original
diagram would be one-pair reducible, not a skeleton as assumed.

It follows that each $n_c$, if not zero, must be equal to $n$ and the sole
combinations of products allowed are those in Eq. (\ref{apx1}). In practice
the full result applies nontrivially only for $n=2$; see Fig. \ref{F2}.
For higher order at most one of the expectations can survive.

Equation (\ref{apx1}) has the following consequence for expectations
of anticouplings: the Kraichnan average of $\overline\varphi$
for any skeleton graph in the complete Luttinger-Ward functional is
\begin{eqnarray}
{\Big\langle \prod^n_{i=1} (1 - \varphi_i) \Big\rangle}_K \!\!
&=&
1 - {\Big\langle \prod^n_{i=1} \varphi_i \Big\rangle}_K.
\label{apx3.0}
\end{eqnarray}
If the left-hand side of Eq. (\ref{apx3.0}) vanishes, it can only be
when the expectation on the right is unity, since the skeleton must be
of $stu$ form. On the the other hand, if the left-hand side does not
vanish, it must be unity while, on the right-hand side, the expectation
vanishes if and only if
\begin{eqnarray*}
0 = {\Big\langle \prod^n_{i=1} s_i \Big\rangle}_K
  = {\Big\langle \prod^n_{i=1} t_i \Big\rangle}_K
  = {\Big\langle \prod^n_{i=1} t_i \Big\rangle}_K
\end{eqnarray*}
or, in other words, when the skeleton is not in the $stu$ set.

\section{Extraction of primitively irreducible kernel}

Our premise is that there exists a well defined kernel $\widehat \Xi$
generating the complementary non-$stu$ kernel $\Xi$, defined in
Eq. (\ref{irrX1.0}), with which all its pair-reducible components
can be obtained. We assume the parquet-like equations

\begin{eqnarray}
\Xi_s
&=&
{\widehat \Xi} + {\overline\phi}^{-1} (
  \Xi\ophi :GG: \ot\Xi_t
- \Xi\ophi :GG: \ou\Xi_u
);
\cr
\Xi_t
&=&
{\widehat \Xi} + {\overline\phi}^{-1} (
  \Xi\ophi :GG: \os\Xi_s
- \Xi\ophi :GG: \ou\Xi_u
);
\cr
\Xi_u
&=&
{\widehat \Xi} + {\overline\phi}^{-1} (
  \Xi\ophi :GG: \os\Xi_s
+ \Xi\ophi :GG: \ot\Xi_t
)
\cr
\cr
{\rm with}~~
\Xi
&\equiv&
{\widehat \Xi} + {\overline\phi}^{-1} (
  \Xi\ophi :GG: \os\Xi_s
+ \Xi\ophi :GG: \ot\Xi_t
~~~ ~~~ ~~~ ~~~ 
\cr
&&
-~ \Xi\ophi :GG: \ou\Xi_u
).
\label{apx3}
\end{eqnarray}

The last expression in Eq. (\ref{apx3}) can also be cast as
\begin{eqnarray}
\Xi
&=&
\Xi_s + {\overline\phi}^{-1} \Xi\ophi :GG: \os\Xi_s
\cr
&=&
\Xi_t + {\overline\phi}^{-1} \Xi\ophi :GG: \ot\Xi_t
\cr
&=&
\Xi_u - {\overline\phi}^{-1} \Xi\ophi :GG: \ou\Xi_u
\label{apx4}
\end{eqnarray}
leading to the formal solutions for the auxiliary kernels
\begin{eqnarray}
\Xi_s
&=&
(II + {\overline\phi}^{-1} \Xi\ophi :GG: \os)^{-1}\Xi;
\cr   
\Xi_t
&=&
(II + {\overline\phi}^{-1} \Xi\ophi :GG: \ot)^{-1}\Xi;
\cr   
\Xi_u
&=&
(II - {\overline\phi}^{-1} \Xi\ophi :GG: \ou)^{-1}\Xi.
\label{apx5}
\end{eqnarray}
Now from Eqs. (\ref{apx3}) through (\ref{apx5}) we can also rewrite $\Xi$ as
\begin{eqnarray*}
\Xi
&=&
\tfrac{1}{2}(\Xi_s + \Xi_t + \Xi_u - \widehat\Xi)
\end{eqnarray*}
to arrive at a final formulation for $\widehat\Xi$ purely in terms of $\Xi$
and the selective channel anticoupling factors:
\begin{widetext}
\begin{eqnarray}
\widehat\Xi
&=&
{\Bigl[
  (II + {\overline\phi}^{-1} \Xi\ophi :GG: \os)^{-1}
+ (II + {\overline\phi}^{-1} \Xi\ophi :GG: \ot)^{-1}
+ (II - {\overline\phi}^{-1} \Xi\ophi :GG: \ou)^{-1}
\Bigr]} \Xi - 2\Xi
\label{apx7}
\end{eqnarray}
rolling back, in each channel, all the pair-reducible iterations of
$\widehat \Xi$ within $\Xi$. The kernel thus obtained is closely related
to the complete, primitively irreducible kernel invoked in parquet theory.

Although from the Kraichnan viewpoint there is no real relevance to
Eq. (\ref{apx7}), rewriting the exact Eqs. (\ref{kII15.A}) and
(\ref{kII15.B}) in terms of $\widehat \Xi$ results in a set of
equations more closely resembling classical parquet, namely
\begin{eqnarray}
\Gamma_s
&=&
2\OV +
{\widehat \Xi}[\OV \ophi] + ({\phi \!+\! \overline \phi})^{-1} {\Bigl(
  \Gamma(\varphi \!+\! \ophi) :GG: (t \!+\! \ot)\Gamma_t
- \Gamma(\varphi \!+\! \ophi) :GG: (u \!+\! \ou)\Gamma_u
\Bigr)};
\cr
\Gamma_t
&=&
\OV +
{\widehat \Xi}[\OV \ophi] + ({\phi \!+\! \overline \phi})^{-1} {\Bigl(
- \Gamma(\varphi + \ophi)\varphi :GG: (u \!+\! \ou)\Gamma_u
+ \Gamma(\varphi + \ophi)\varphi :GG: (s \!+\! \os)\Gamma_s
\Bigr)};
\cr
\Gamma_u
&=&
\OV +
{\widehat \Xi}[\OV \ophi] + ({\phi \!+\! \overline \phi})^{-1} {\Bigl(
  \Gamma(\varphi + \ophi) :GG: (s \!+\! \os)\Gamma_s
+ \Gamma(\varphi + \ophi) :GG: (t \!+\! \ot)\Gamma_t
\Bigr)}
\cr
\cr
{\rm for}~~
\Gamma
&\equiv&
2\OV + {\widehat \Xi}[\OV \ophi] 
+ ({\phi \!+\! \overline \phi})^{-1} {\Bigl(
  \Gamma(\varphi \!+\! \ophi) :GG: (s \!+\! \os)\Gamma_s
+ \Gamma(\varphi \!+\! \ophi) :GG: (t \!+\! \ot)\Gamma_t
- \Gamma(\varphi \!+\! \ophi) :GG: (u \!+\! \ou)\Gamma_u
\Bigl)}.
~~~ ~~~ ~~~ ~~~ 
\label{apx8}
\end{eqnarray}
\end{widetext}
This version of exact parquet departs in two significant ways from the
standard case. First, the K coupling compensates for overcounting when
$\Phi$ is reconstructed from Eq. (\ref{apx8}). This correction can also
be applied, if by hand, in the context of normal parquet. Second,
to represent systematically all the structural combinations in the exact
$\Gamma$, the K couplings and their anticouplings operate independently of
one another, despite the fact that they sum identically to unity.
In that way the variationally defined kernel stays isomorphic with
the structure of the generating Luttinger-Ward functional.
\\


\begin{thebibliography}{99}

\bibitem{becker}
W. Becker and D. Grosser, Nuov. Cim. A {\bf 10}, 343 (1972).

\bibitem{js} 
A. D. Jackson and R. A. Smith,
Phys. Rev. A {\bf 36}, 2517 (1987).

\bibitem{roger} 
R. A. Smith,
Phys. Rev. A {\bf 46}, 4586 (1992).

\bibitem{lw} 
J. M. Luttinger and J. C. Ward,
Phys. Rev. {\bf 118}, 1417 (1960).

\bibitem{kb1} 
G. Baym and L. P. Kadanoff,
Phys. Rev. {\bf 124}, 287 (1961).

\bibitem{kb2} 
G. Baym,
Phys. Rev. {\bf 127}, 1391 (1962).

\bibitem{pqt1} 
R. W. Haymaker and R. Blankenbecler, 
Phys. Rev. {\bf 171}, 1581 (1968).

\bibitem{pqt2} 
A. D. Jackson, A. Lande, and R. A. Smith,
Phys. Rep. {\bf 86}, 55 (1982).

\bibitem{pqt2a} 
A. Lande, and R. A. Smith,
Phys. Lett. B {\bf 131}, 253 (1983).

\bibitem{pqt3} 
N. E. Bickers,
Int. J. Mod. Phys. B {\bf 5}, 253 (1991).

\bibitem{k1} 
R. H. Kraichnan, 
J. Math. Phys. {\bf 3}, 475 (1962).

\bibitem{k2} 
R. H. Kraichnan, 
J. Math. Phys. {\bf 3}, 496 (1962).

\bibitem{KI} 
F. Green,
Phys. Rev. A {\bf 99}, 062118 (2019). 

\bibitem{dp} 
D. Pines and P. Nozi\`eres,
{\em The Theory of Quantum Liquids, Volume I:
Normal Fermi Liquids}
(Benjamin, New York, 1966).

\bibitem{pn} 
P. Nozi\`eres, {\em Theory of Interacting Fermi Systems}
(Benjamin, New York, 1964), Ch. 5.

\bibitem{KII} 
F. Green and T. L. Ainsworth,
Phys. Rev. A {\bf 106}, 052208 (2022). 

\bibitem{sb} 
N. E. Bickers, D. J. Scalapino, and S. R. White,
Phys. Rev. Lett. {\bf 62}, 961 (1989).

\bibitem{ddm}
C. De Dominicis and P. C. Martin,
J. Math. Phys. {\bf 5}, 14 (1964).

\bibitem{kita}
T. Kita,
J. Phys. Soc. Jpn. {\bf 91}, 114002 (2022).

\bibitem{potthoff}
M. Potthoff, Condens. Matter Physics {\bf 9}, 557 (2006);
https://doi.org/10.48550/arXiv.cond-mat/0406671.

\bibitem{lin1} 
L. Lin and M. Lindsey,
Proc. Natl. Acad. Sci. {\bf 115}, 2282 (2018);
Arch. Rational Mech. Anal. {\bf 242}, 581 (2021).

\bibitem{lin2} 
L. Lin and M. Lindsey,
Arch. Rational Mech. Anal. {\bf 242}, 527 (2021).

\bibitem{cons}
Conservation of the collective-index sum comes from invariance of the
individual embedded Hamiltonians, analogous to translational
invariance of the individual unit cells embedded in a lattice, leading to
a conserved crystal momentum.

\bibitem{xsym}
For the three auxiliary $stu$ kernels, $\Lambda_s$ is antisymmetric
while $\Lambda_t \rightleftharpoons -\Lambda_u$ on label exchange.
Dropping $\Lambda''$ in the $stu$ form of the parquet equations
results in a total kernel that is crossing symmetric though
no longer conserving. It is the analog of minimal standard parquet
\cite{pqt3}
where $\OV $ is the only irreducible input kernel.


\bibitem{qsm}
G. Baym,
in {\em Progress in Nonequilibrium Green's Functions},
M. Bonitz ed. (World Scientific, Singapore, 2000), pp 17-32.

\bibitem{tom}
T. L. Ainsworth and K. S. Bedell,
Phys. Rev. B {\bf 35}, 8425 (1987).

\bibitem{vll}
V. Llisie, {\em Concepts in Quantum Field Theory}
(Springer, Cham, 2016), Ch. 12. 

\bibitem{mahan2}
G. D. Mahan, {\em Many-Particle Physics} (Plenum, New York, 1981),
Ch. 5.

\bibitem{a+m}
N. W. Ashcroft and N. D. Mermin, {\em Solid state physics}
(Saunders, New York, 1976). 

\bibitem{dmft}
A. Georges, G. Kotliar, W. Krauth, and M. Rozenberg, {\em Rev. Mod. Phys.}
{\bf 68}, 13 (1996).

\bibitem{lande}
A. Lande and R. A. Smith, Phys. Rev. A {\bf 45}, 913 (1992).

\end{thebibliography}
\end{document}